\documentclass[sigconf]{acmart}
\usepackage{makecell}
\usepackage{multirow}
\usepackage{colortbl}
\usepackage{subcaption} 
\usepackage{enumitem}

\AtBeginDocument{%
  }

\copyrightyear{2026}
\acmYear{2026}
\setcopyright{cc}
\setcctype{by-nc-nd}
\acmConference[C\&C '26]{Creativity and Cognition}{July 13--16, 2026}{London, United Kingdom}
\acmBooktitle{Creativity and Cognition (C\&C '26), July 13--16, 2026, London, United Kingdom}
\acmDOI{10.1145/3803784.3807536}
\acmISBN{979-8-4007-2583-8/2026/07}

\begin{document}

\title[What Makes an AI Writing Companion a Good Fit?]{What Makes an AI Writing Companion a Good Fit? A Personality-Informed Co-Design Study}

\author{Mengke Wu}
\email{mengkew2@illinois.edu}
\affiliation{
    \institution{University of Illinois Urbana-Champaign}
    \department{School of Information Sciences}
    \city{Champaign}
    \state{Illinois}
    \country{USA}
}

\author{Kexin Quan}
\email{kq4@illinois.edu}
\affiliation{
    \institution{University of Illinois Urbana-Champaign}
    \department{School of Information Sciences}
    \city{Champaign}
    \state{Illinois}
    \country{USA}
}

\author{Weizi Liu}
\email{weizi.liu@tcu.edu}
\affiliation{
    \institution{Texas Christian University}
    \department{Bob Schieffer College of Communication}
    \city{Fort Worth}
    \state{Texas}
    \country{USA}
}

\author{Mike Yao}
\email{mzyao@illinois.edu}
\affiliation{
    \institution{University of Illinois Urbana-Champaign}
    \department{Institute of Communications Research}
    \city{Champaign}
    \state{Illinois}
    \country{USA}
}

\author{Jessie Chin}
\email{chin5@illinois.edu}
\affiliation{
    \institution{University of Illinois Urbana-Champaign}
    \department{School of Information Sciences}
    \city{Champaign}
    \state{Illinois}
    \country{USA}
}

\renewcommand{\shortauthors}{Wu et al.}

\begin{abstract}
The growing popularity of AI writing assistants creates exciting opportunities to support diverse writers. This study examines how personality shapes expectations for AI writing companions and how personality-informed design can enhance human–AI teaming in writing. Through exploratory co-design workshops with 24 writers representing different personality profiles, we elicited values and design ideas for AI writing companions spanning functionality, interaction dynamics, and visual representation. These insights informed two contrasting prototypes reflecting distinct writing orientations, used as design provocations in review-and-refinement workshops with eight participants to prompt reflection on fit, priorities, and writing practices. Our findings reveal both shared foundational needs across writers and meaningful personality-driven preferences that influence how writers engage with AI. This work underscores the importance of team matching in human-AI collaboration and demonstrates how aligning AI companions with individual cognitive and interpersonal needs can improve engagement and perceived collaboration effectiveness.
\end{abstract}

\ccsdesc[500]{Human-centered computing~Empirical studies in HCI}
\ccsdesc[500]{Human-centered computing~User studies}
\ccsdesc[500]{Human-centered computing~Participatory Design}
\ccsdesc[300]{Human-centered computing~Natural language interfaces}
\ccsdesc[300]{Human-centered computing~Collaborative and social computing design and evaluation methods}

\keywords{Co-design, Chatbot, Personality, Human-AI teaming, Creativity support, User-centered design}


\maketitle

\section{INTRODUCTION}

Writing is a deeply personal and cognitively demanding activity that varies in how individuals generate ideas, structure content, reflect on their work, and manifest creativity \cite{levy2013science, kufner2010tell}. Prior research has shown that personality plays a significant role in shaping these practices, influencing stylistic choices, revision strategies, and feedback interpretation \cite{jensen1984personality, pennebaker1999linguistic, kufner2010tell}. For example, writers with analytical or conscientious tendencies often favor structured, detail-oriented compositions, whereas those with intuitive or empathetic orientations prioritize flexibility, creativity, and emotional tone {\cite {kufner2010tell, wolfradt2001individual, jensen1984personality}}. These differences suggest that effective writing support depends not only on correctness or efficiency, but also on cognitive and interpersonal resonance \cite{zhang2021ideal, rezwana2022understanding, hemmer2025complementarity}.

\begin{figure*}[t]
    \centering
    \includegraphics[width=1\linewidth]{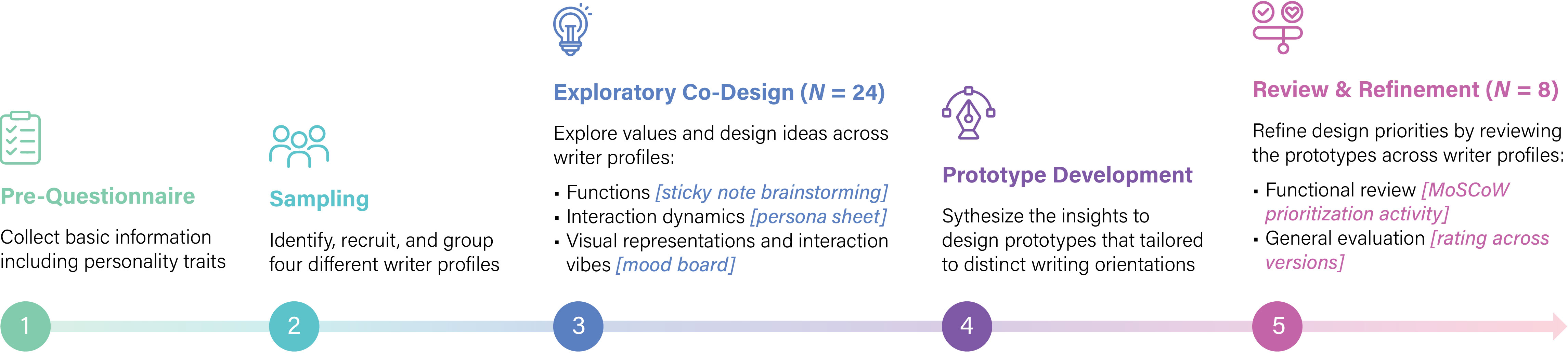}
    \caption{Overall Project Workflow: From Writer Profiling to Co-Design and Review.}
    \Description{A five-step horizontal workflow diagram showing the study process: (1) Pre-Questionnaire collecting personality traits; (2) Sampling identifying four writer profiles; (3) Exploratory Co-Design (N=24) eliciting design ideas through brainstorming, persona sheets, and mood boards; (4) Prototype Development synthesizing insights into two contrasting prototypes; (5) Review and Refinement (N=8) refining design priorities through MoSCoW prioritization and prototype ratings.}
    \label{fig:workflow}
\end{figure*}

Recent advances in large language models (LLMs), such as ChatGPT\footnote{https://chat.openai.com/} and Grammarly\footnote{https://www.grammarly.com/}, have enabled AI writing assistants capable of supporting tasks ranging from grammar correction and stylistic refinement to ideation and content generation \cite{gero2022sparks, yuan2022wordcraft, clark2018creative, dang2022beyond}. While widely adopted, these systems do not necessarily support all writers equally well, particularly in tasks closely tied to identity, creativity, and a sense of ownership \cite{bhat2023approach, nicholes2017measuring}. Research has increasingly explored human-AI collaboration in writing contexts, revealing that users hold varied expectations of AI writing systems, viewing them as collaborators, editors, or utilitarian tools \cite{ding2023mapping, yang2022ai, biermann2022tool}. Personality has emerged as an important factor underlying these differences, as writers bring distinct goals, habits, and interaction preferences into co-creative writing scenarios \cite{jakesch2023co, biermann2022tool, gero2019metaphoria}. Understanding these individual differences is critical: misalignment between user expectations and system behavior can affect collaboration quality, trust, and long-term engagement \cite{zhang2021ideal, huang2019human, bach2024systematic}. This tension motivates how we design for diverse user needs.

While personalization has been an important focus in AI writing research, existing efforts often emphasize internal algorithmic adaptation (e.g., \cite{tang2024step, gabriel2015inkwell}), which primarily target output customization rather than the collaborative relationship itself. Some studies have explored interface-level customization and controllable features to tailor writing processes and outcomes (e.g., \cite{yeh2024ghostwriter, yuan2022wordcraft, shi2022effidit}), but often adopt a one-size-fits-all approach and involve limited user participation in design processes. Rather than asking whether users will adopt AI writing tools, a more critical question is how well these systems fit the diverse ways writers think, create, and collaborate.

Our research investigates personality-driven design for AI writing companions by asking: \textbf{How do personalities shape user preferences for AI writing companions, and what design priorities emerge?} We formulated two design questions:

\begin{itemize}
\item DQ1: How do writers with different personality profiles envision desirable functions, interaction dynamics, and visual representations for writing companions?
\item DQ2: Which characteristics receive consistent endorsement within each profile, and how do these patterns refine priorities for personality-aware design?
\end{itemize}

To address these questions, we conducted a multi-phase study (see Figure \ref{fig:workflow}). We first administered a pre-questionnaire with Big Five \cite{rammstedt2007measuring} to characterize personality traits within a larger screening pool. We focused on dimensions closely linked to creative, cognitive, and interpersonal orientations (i.e., Openness and Agreeableness) \cite{kaufman2016openness, upadhyaya2023bot, pennebaker1999linguistic, wolfradt2001individual} to recruited representative writers. In \textbf{exploratory co-design workshops (\textbf{DQ1})}, 24 participants elicited values, expectations, and design ideas for AI writing companions' functions, interaction dynamics, and visual representations. These insights informed two contrasting personality-informed prototypes. We then conducted \textbf{review-and-refinement workshops (\textbf{DQ2})} with eight participants, using these prototypes as design provocations to stimulate reflection and discussion about fit and priorities, thereby refining the emerging design space.

Through this process, our work moves beyond one-size-fits-all approaches to AI writing tools and contributes:
\begin{itemize}
    \item Empirical insights into how personality profiles articulate preferences for AI-supported writing.
    \item A personality-informed design space grounded in participatory input and iterative prototyping.
    \item Actionable implications for designing more inclusive and supportive AI writing companions.
\end{itemize}

\section{RELATED WORK}

\subsection{Individual Differences in Writing}\label{individual}
Writers differ in how they express themselves, shaped by multiple factors. For example, cognitive styles affect planning and revision strategies \cite{flower1981cognitive, mccutchen2000knowledge}, sociocultural backgrounds shape rhetorical structure and cohesion \cite{kaplan1966cultural, 
silva1993toward}, and communication styles leave a mark on written tone and structure \cite{ivanov2010behavioral}.

Among these factors, personality plays a particularly influential role. Defined as stable patterns of thinking, feeling, and behaving across time and contexts \cite{funder2012accurate}, personality affects not only social interactions but also leaves measurable traces in linguistic expression and writing behavior. For instance, extraverted and agreeable individuals use more social and positive-emotion words, and those high in openness display richer vocabulary and creativity \cite{pennebaker1999linguistic, kufner2010tell, wolfradt2001individual}. Beyond stylistic markers, personality influences cognitive and motivational processes underlying writing tasks, shaping how writers approach revision, interpret feedback, tolerate ambiguity, and manage emotions \cite{jensen1984personality, Banaruee2017CorrectiveFA, Marefat2006STUDENTWP}. Some prefer exploratory feedback that encourages creative risk-taking, while others favor specific, directive feedback focused on correctness and structure \cite{perks2021role, yunita2022students}. Writers also differ in their comfort with open-ended tasks, desire for control, and sensitivity to ownership \cite{biermann2022tool, reza2025co}. Their approaches to collaboration vary as well: some embrace interaction and negotiation as integral to composing and refining ideas, whereas others prefer to treat writing as a solitary activity with clear individual control over the text \cite{reza2025co, savacscci2019one, wang2017users}. These differences influence how writers engage with peers, instructors, or AI systems, suggesting that writing is not merely a skill but a deeply personal activity shaped by individual characteristics.
Yet these personality-driven differences have rarely been translated into concrete design principles for AI writing systems, a gap this study directly addresses.

\subsection{Human-AI Teaming in Writing}
Advances in AI expand opportunities for collaborative human–AI interaction, enabling systems to act as both assistants and co-cognitive partners for improved outcomes \cite{park2019identifying, sowa2021cobots, wang2020human, kim2022learning, quan2026towards}. Effective human-AI teaming depends on shared goals, aligned mental models, and coordinated communication styles to support trust, decision-making, and team performance \cite{zhang2021ideal, huang2019human, liang2019implicit}.  Writing has become a prominent domain for such collaboration, with tools like ChatGPT widely adopted as writing assistants\footnote{https://sparktoro.com/blog/we-analyzed-millions-of-chatgpt-user-sessions-visits-are-down-29-since-may-programming-assistance-is-30-of-use/} \footnote{https://www.statista.com/statistics/1378998/chatgpt-use-tasks-us-by-type/}. This reliance motivates our investigation into how human–AI teaming can be designed to better support co-writing practices.

\subsubsection{AI Collaboration in Writing: Divergence and Tensions}
Research on AI-supported writing reveals complex and sometimes contradictory user experiences. Some studies document enthusiasm for AI as creative partners: users report drawing inspiration from AI-generated text and enjoying collaborative writing, even when AI contributions do not objectively improve output quality \cite{yang2022ai, clark2018creative}. Related work on AI in narrative construction, dialogue generation, and educational feedback further highlights its capacity to meaningfully participate in writing processes \cite{serban2017hierarchical, gao2018neural, kangasharju2022lower, quan2025can}. However, other research surfaces persistent tensions around control, agency, and authorship. Studies found that creative writers, while open to AI assistance, strongly emphasized the need for systems that respect their personal values and writing strategies rather than imposing external voices \cite{biermann2022tool}. This reflects deeper concerns about ownership and emotional attachment writers develop toward their work \cite{furby1978possession, dittmar1991meanings, belk1988possessions, nicholes2017measuring}. They are further amplified by findings that AI can subtly influence users: awareness of AI authorship may reduce trust \cite{liu2022will}, and opinionated language models can shift attitudes on controversial topics \cite{jakesch2023co}. These divergent findings highlight an unresolved tension: AI writing support may not be universally defined but rather dependent on individual differences in goals, values, cognitive orientations, and preferred collaboration styles \cite{biermann2022tool, gero2019metaphoria, reza2025co}.
Our work responds to this by exploring how these differences inform system design, making individual differences a first-class design consideration rather than an afterthought.

\subsubsection{Personalization in AI Writing}
As AI writing tools become embedded in everyday practice, personalization emerges as a key design concern, as tailoring systems to users could address the tensions identified above. This underscores the need for AI writing systems that scaffold individuals’ task preferences while honoring their needs for personal connection and creative autonomy.

Most existing approaches to personalization emphasize back-end adaptation, such as implicitly learning from user inputs to refine text generation \cite{tang2024step, kim2019designing, salemi2023lamp}. For instance, \textit{InkWell} generates stylistic variations by mimicking writers’ tone and personality during revision \cite{gabriel2015inkwell}, while recent work leverages user history profiling to tailor scientific writing assistance \cite{tang2024step}. Beyond algorithmic adaptation, some studies explore front-end design strategies, including domain-specific systems (e.g., language learning \cite{gayed2022exploring}, scientific writing and peer review \cite{kacena2024use, sun2024metawriter}, creative storytelling \cite{yang2022ai}) to broader design efforts aimed at improving usability, transparency, and user control \cite{yeh2024ghostwriter, yuan2022wordcraft, shi2022effidit}. For example, \textit{Wordcraft} supports story co-writing through a dialogue-based interface that allows users to interactively steer model outputs \cite{yuan2022wordcraft}, while \textit{GhostWriter} explores personalization through agency-preserving design patterns \cite{yeh2024ghostwriter}. However, many efforts adopt one-size-fits-all designs and rely on post-hoc usability evaluation of the proposed systems, overlooking individual differences and user-informed opportunities in system design. Recent work in HCI calls for more human-centered, participatory approaches that involve users early in shaping AI systems \cite{fitzsimons2024overcoming, van2024exploring, delgado2023participatory}. Our work builds on these calls by engaging writers in the ideation stage to co-design personalized writing companions that reflect diverse lived experiences, expectations, and collaboration styles.

\section{EXPLORATORY CO-DESIGN WORKSHOP}

\subsection{Participants and Writer Profiles}
Participants were recruited from a U.S. public university via email lists and an online research pool, following IRB Approval. A pre-questionnaire collected demographic information, writing self-efficacy scores \cite{bruning2013examining}, prior design experience, and personality traits using the brief Big Five inventory \cite{rammstedt2007measuring}. The screening survey received responses from 168 individuals, which we used to estimate the relative distribution of personality traits within our recruitment pool and to guide balanced participant selection.

Among the Big Five dimensions, we focused on Openness and Agreeableness, as prior research consistently links them to creativity, writing styles, cognitive exploration, or interpersonal preferences in writing and interaction contexts \cite{kaufman2016openness, pennebaker1999linguistic, upadhyaya2023bot, wolfradt2001individual}. Openness reflects an orientation toward abstract thinking, imagination, and conceptual exploration, whereas lower Openness is associated with preferences for structure, concreteness, and procedural clarity. Agreeableness reflects interpersonal orientation in judgment and decision-making, ranging from logic- and task-focused evaluation to value-driven, empathetic, and relational considerations \cite{de2000big, kaufman2016openness}.

\begin{figure*}
    \centering
    \includegraphics[width=1\linewidth]{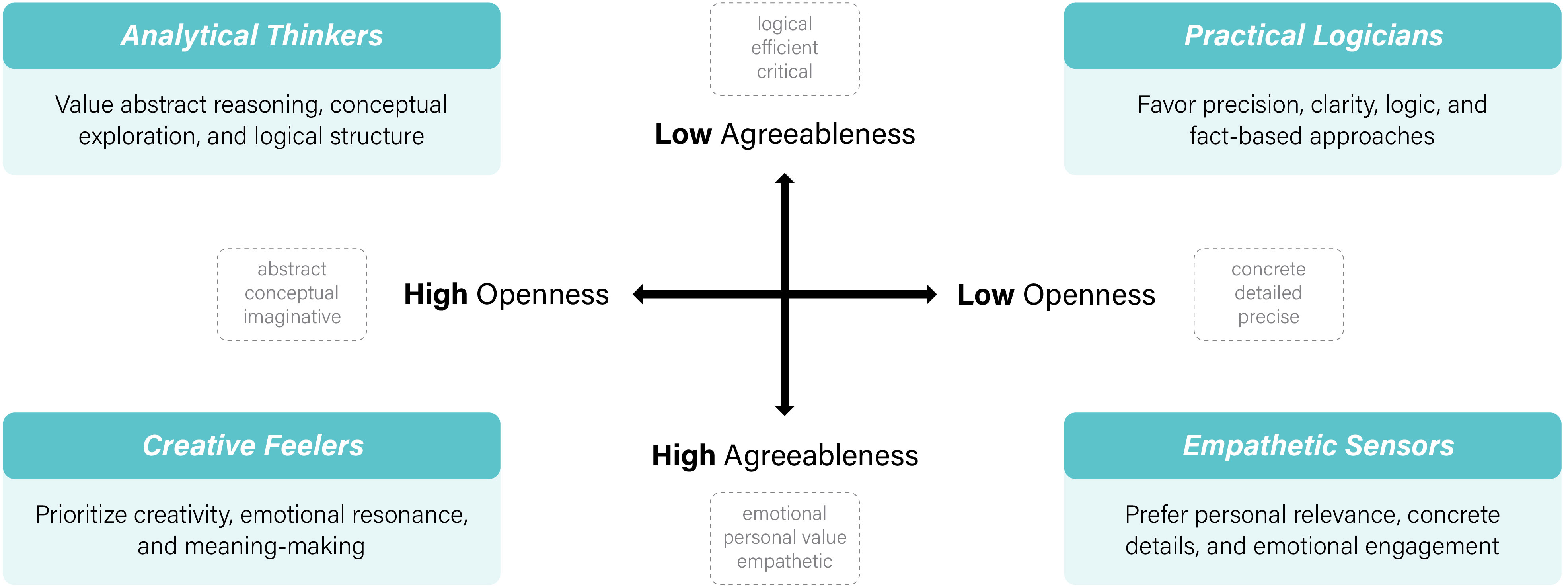}
    \caption{The Four Writer Profiles derived from Personality Traits.}
    \Description{A two-by-two quadrant diagram with Openness on the horizontal axis and Agreeableness on the vertical axis, defining four writer profiles: Analytical Thinkers (high Openness, low Agreeableness) valuing abstract reasoning and logical structure; Practical Logicians (low Openness, low Agreeableness) favoring precision and fact-based approaches; Creative Feelers (high Openness, high Agreeableness) prioritizing creativity and emotional resonance; and Empathetic Sensors (low Openness, high Agreeableness) preferring personal relevance and emotional engagement.}
    \label{fig:profile}
\end{figure*}

Rather than treating personality traits as deterministic labels, we used them as analytic anchors to construct interpretable writer profiles capturing contrasting orientations toward writing and AI collaboration. Using a distribution-based grouping strategy on the full screening sample (\textit{N} = 168), we identified upper and lower quartiles for Openness (O) and Agreeableness (A). Participants within these quartiles were prioritized for recruitment, with secondary consideration given to those near the thresholds to support balanced group sizes. This yielded four writer profiles: 1) \textit{Creative Feelers} (high-O, high-A), who prioritize creativity, emotional resonance, and meaning-making; 2) \textit{Analytical Thinkers} (high-O, low-A), who value abstract reasoning, conceptual exploration, and logical structures; 3) \textit{Empathetic Sensors} (low-O, high-A), who prefer concrete details grounded in personal relevance and emotional engagement, and 4) \textit{Practical Logicians} (low-O, low-A), who favor precision, clarity, and fact-based approaches (see Figure \ref{fig:profile}). Extraversion, Conscientiousness, and Neuroticism were not used as they primarily reflect social energy, self-regulation, and emotional stability \cite{de2000big}, which are less directly related to how writers conceptualize ideas, interpret feedback, and negotiate interaction styles with AI.

\begin{figure*}
    \centering
    \includegraphics[width=1\linewidth]{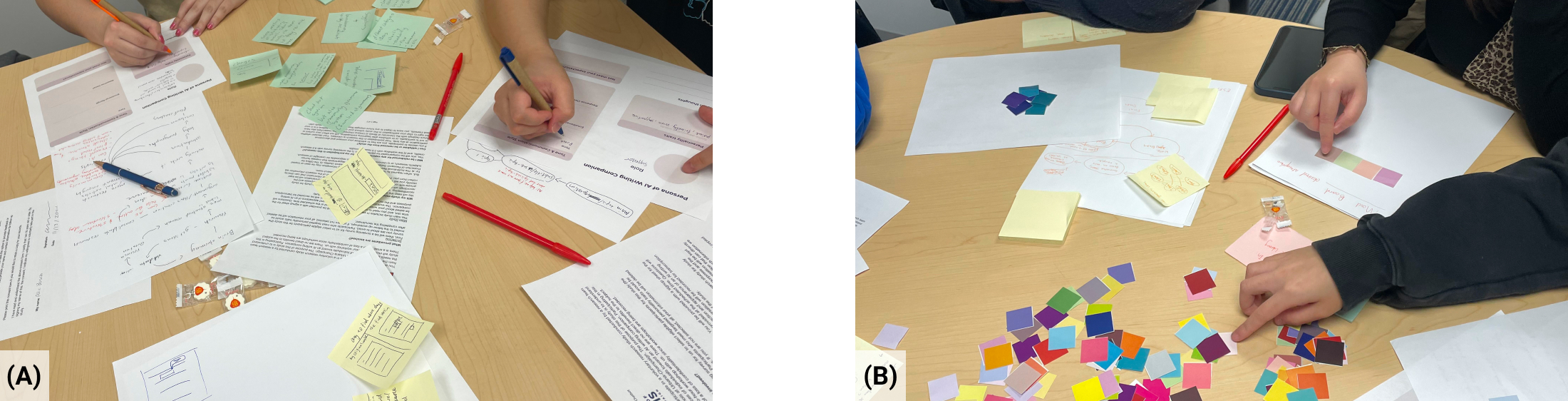}
    \caption{Example Workshop Activities: (A) Brainstorming for Desired Functions and AI Personas, (B) Mood Board Creation.}
    \Description{Two photographs of in-person workshop sessions. Image A shows participants brainstorming around a table covered with sticky notes and worksheets for desired AI companion functions and personas. Image B shows participants creating mood boards with colorful paper pieces to express their envisioned interaction styles with an AI writing companion.}
    \label{fig:photo}
\end{figure*}

In total, 24 participants were recruited (six per profile) and grouped for workshops. Participants had an average age of 23.9 (\textit{SD} = 3.84), with 10 males, 13 females, and one genderqueer. Writing self-efficacy varied widely (\textit{M} = 72.9/100, \textit{SD} = 17.67). Eight participants had design experience, four had some, and nine had none. The sample exhibited clear separation along Openness and Agreeableness in intended directions. Other Big Five dimensions, along with availability and demographics, were jointly considered to reduce potential covariates; however, given practical recruitment constraints, some residual variation remained (see Appendix \ref{appendix:stats} for sampling statistics and Appendix \ref{appendix:profile} for participant profiles).

\subsection{Workshop Materials and Procedure}
We invited participants to co-explore the design space of AI writing companions, eliciting desired functions, traits, and visual representations. Each 75-minute in-person workshops were structured to balance individual reflection with group discussion. The workshop began with an overview of research objectives and session structure, followed by a discussion of participants’ general writing experiences (e.g., what constitutes good writing, challenges in writing). Participants then created flowcharts to illustrate their writing processes and identify where and how generative AI (GAI) currently supported their work. Next, participants engaged in a feature brainstorming activity, recording ideas on sticky notes that included both improvements to existing tools and entirely new capabilities. These ideas were then shared in the group to identify common themes and develop new concepts. Participants then created persona sheets for their ideal companions, specifying preferred roles, personality traits, communication styles (tone, emotional range, and response detail), and how they should handle inaccuracies or misaligned suggestions (see Figure \ref{fig:photo}-A). Finally, participants described their ideal companion’s appearance (e.g., human, animal, object) and created mood boards \cite{lucero2012framing} to visually express their envisioned interaction vibes with the companion (see Figure \ref{fig:photo}-B).

\section{EXPLORATORY DESIGN FINDINGS (DQ1)}\label{ideation}

The co-design workshops were audio-recorded and transcribed with participants' consent, capturing discussions, design rationales, and reflections throughout ideation. Workshop artifacts (e.g., sketches, written sheets, mood boards) were also collected and analyzed as complementary data in the thematic analysis \cite{braun2024thematic}. Two researchers independently coded the data, reconciled differences, and iteratively refined the codebook. The final code structure followed the workshop activities, including writing styles and preferences, desired features, interaction dynamics, and visual representations. Sub-themes emerged from these categories, such as \textit{“Writing/Response Style”} under desired features. Smaller topics, like \textit{"Creative Feelers: emotional appeal,"} were grouped as profile-specific patterns.

\subsection{Desired Functionalities for AI Writing Companions}

\begin{figure*}
    \centering
    \includegraphics[width=1\linewidth]{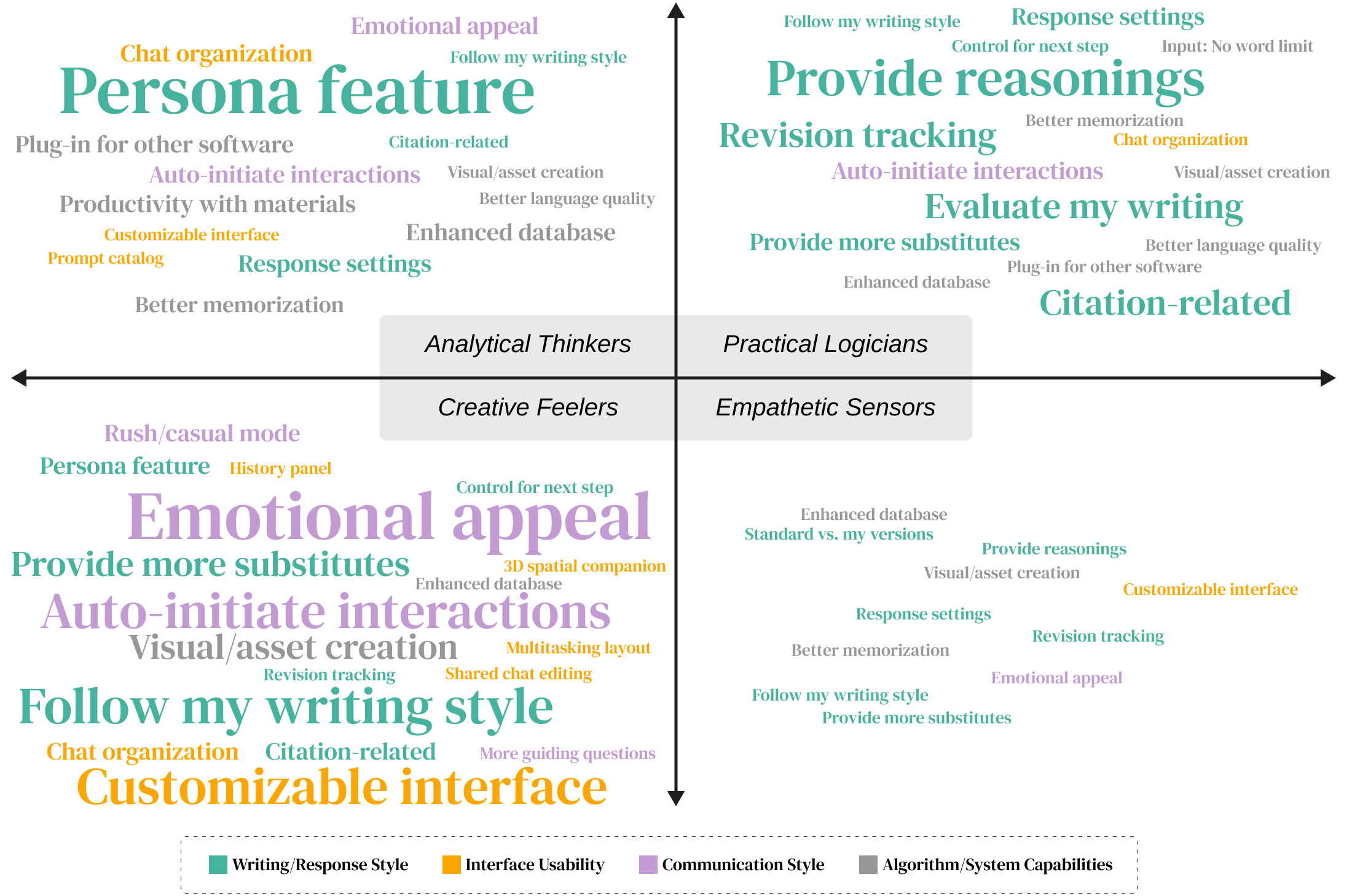}
    \caption{Frequency-Weighted Word Cloud of Proposed Features by Different Writer Profiles.}
    \Description{Four word clouds, one per writer profile, where word size indicates frequency of participant mentions, color-coded by category: Writing/Response Style, Interface Usability, Communication Style, and Algorithm/System Capabilities. Prominent features include "Persona feature" and "Emotional appeal" for Analytical Thinkers; "Provide reasonings" and "Revision tracking" for Practical Logicians; "Customizable interface" and "Follow my writing style" for Creative Feelers; and "Provide reasonings" and "Customizable interface" for Empathetic Sensors.}
    \label{fig:wordcloud}
\end{figure*}

This section outlines functional design space for AI writing companions. As an illustrative overview, Figure \ref{fig:wordcloud} visualizes the features proposed across writer profiles, with prominence indicating the frequency of participant mentions.

\subsubsection{Writing/Response Style} 
A central theme across participants (7 of 24) was the desire for the AI companion to \textbf{adapt to their individual writing styles} by remembering past interactions, with suggestions including a slider to control how much the AI retains of their personal style (P9) and the ability to receive dual outputs: one standard AI version and one reflecting the user’s own style (P15). Building on this, a \textbf{persona feature} was particularly prominent among five high-O participants (\textit{Analytical Thinkers} \& \textit{Creative Feelers}), reflecting their conceptual thinking processes. They felt that existing tools (e.g., ChatGPT) inadequately support this need through static preference settings. P8 suggested an onboarding survey to help the AI learn about the user upfront, while P7 proposed a \textbf{switchable persona feature} that allows context-dependent profiles, such as "student" for precision and clarity, "project manager" for business-oriented communication, or "club organizer" for creative, engaging language. P3 extended this idea conversely, proposing audience-facing personas to adapt language for different readers (e.g., simplify language or use slang for Gen Z).

\begin{figure*}
    \centering
    \includegraphics[width=1\linewidth]{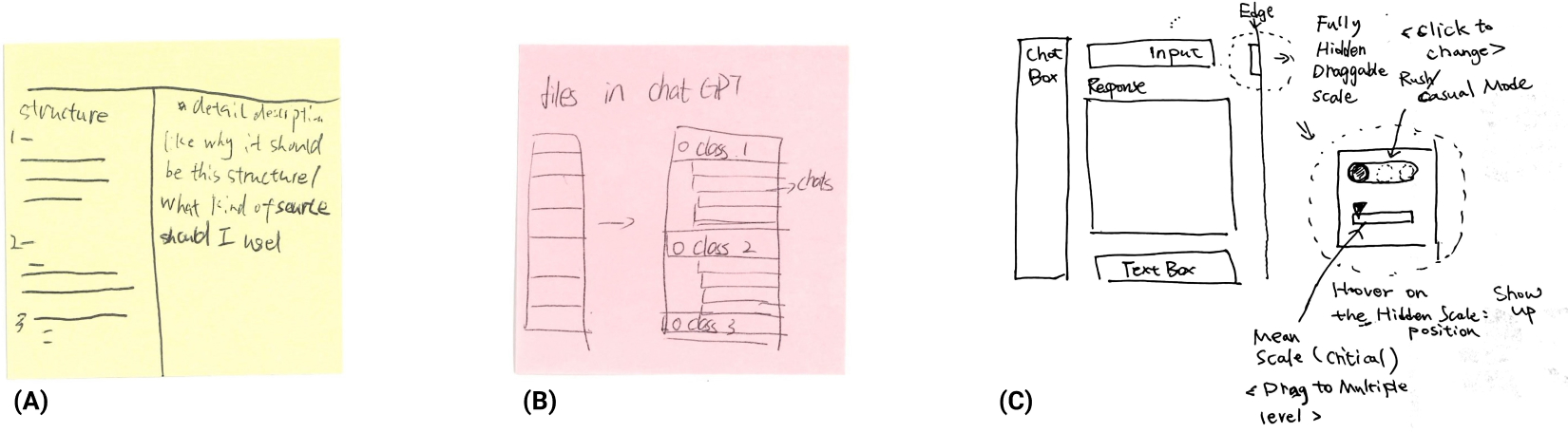}
    \caption{Example Design Sketches from Participants during the Workshop: (A) Providing Reasoning Side-by-Side, (B) Chat Organization, (C) Emotional Scale and Rush/Casual Mode.}
    \Description{Three hand-drawn sketches from workshop participants. Sketch A shows a side-by-side panel displaying AI revision reasoning alongside original text. Sketch B illustrates a chat organization interface grouping conversations into labeled collections. Sketch C depicts an emotional scale and rush/casual mode control panel with sliders to adjust the companion's tone and interaction style.}
    \label{fig:draw}
\end{figure*}

For low-O and low-A profiles (\textit{Practical Logicians} \& \textit{Empathetic Sensors} \& \textit{Analytical Thinkers}), \textbf{response customization} was a priority. Four prioritized fine-grained control over AI outputs, such as tone (formal to casual), focus (e.g., clarity, grammar), context (e.g., academic, creative), and format (bullet points or paragraphs). When requesting revisions, detail-oriented participants (\textit{Practical Logicians} \& \textit{Empathetic Sensors}) highlighted better \textbf{revision tracking}. P22 and P23 praised Grammarly’s approach to highlighting changes, with P23 proposing a toggle to bold/unbold changes. P15 and P23 further recommended side-by-side comparisons or clickable controls to switch between original and revised versions. Transparency around revisions was also important to the low-O profile, with four requesting \textbf{rationales for AI revisions}. Layout ideas included a side-by-side panel (see Figure \ref{fig:draw}-A), end-of-text explanation list, hover-based rationale review, and a toggle to show/hide reasons. Notably, a feature cut across all profiles: \textbf{flexibility in receiving revision suggestions}. Seven participants desired the ability to request and review multiple alternatives for words or phrases in the given outputs (e.g., highlight and click to request).

Smaller but noteworthy features also emerged. Participants raised the need related to \textbf{citations and credibility}. \textit{Practical Logicians} (P20, P24) advocated for auto-generated citation lists with the ability to toggle them on or off. In contrast, \textit{Creative Feelers} (P4, P5) emphasized distinguishing trustworthy sources from AI-generated content, proposing visible “verified” labels to support credibility judgments. Additionally, the fact-driven \textit{Practical Logicians} (P21, P22) expressed interest in receiving \textbf{evaluative feedback}, such as numerical grades, reading levels, and reading time.

\subsubsection{Interface Usability} This theme focuses on interface and visual elements, dominantly articulated by high-O and high-A profiles (\textit{Analytical Thinkers} \& \textit{Creative Feelers} \& \textit{Empathetic Sensors}). Six of them emphasized a \textbf{customizable interface} to match cognitive and emotional preferences, such as altering colors and backgrounds for specific tasks or moods. Three \textit{Creative Feelers} (P4, P5, P6) further envisioned task-based (e.g., plain for coding, vibrant for creative writing, and beach for travel itineraries) or time-based variations (e.g., “sky blue” in the morning, “orange” at lunchtime) to reflect shifts in energy and focus. P1 also re-imagined a multitasking layout, dividing the interface into blocks for the main task, thought trees, checklists, source lists, conversation with the companion, and a history panel, thus streamlining task organization and reducing cognitive load. The \textbf{history panel}, envisioned like Photoshop, offers a navigational workflow overview while preserving change records for back-and-forth experimentation across steps.

Another key preference, particularly among the high-O profile, was \textbf{chat organization} (see Figure \ref{fig:draw}-B). P4 and P11 expressed a desire to group related chats or organize them into custom collections, such as by project, topic, or focus. Although only one \textit{Practical Logician} (P19) commented on chat organization, his suggestion stood out: an auto-generated \textbf{table of contents} with editable anchor points (similar to Google Docs) to improve within-chat navigation and topic tracking for users who value precision and efficiency. Beyond chat, P7 also noted the need to organize prompts. She proposed a \textbf{“prompt catalog”}, as a library, to save or favorite frequently used prompts, streamlining repetitive typing tasks and prompt retrieval.

\subsubsection{Communication Style} Another popular topic was communication style, including the companion's conversational tone and dialogue structure during interaction. Insights from high-O and high-A profiles centered on \textbf{emotional appeal} (\textit{Creative Feelers} \& \textit{Empathetic Sensors}), with three suggesting the AI’s tone could be enriched with greater emotion and appreciating anthropomorphic elements. However, \textit{Creative Feelers} also articulated reservations about overly friendly communication. P5 explained, \textit{“I don't think the seemingly friendly tone really helps users improve their writing skills. Instead, it just violently changes [the writing] to a certain style.”} In response, three high-As (P5, P6, P13) proposed a \textbf{"dragging emotional scale"} to adjust the companion’s critique level based on user needs (e.g., friendly to harsh) (see Figure \ref{fig:draw}-C). P6 also suggested an option to disable emotional elements entirely for users in a rush.

\textit{Creative Feelers} also provided unique input on dialogue style. For example, P5 wanted \textbf{more guiding questions} to encourage cognitive engagement, noting that constantly and merely receiving answers \textit{“won’t become a knowledge of mine.”} Following this, P5 and P6 proposed a \textbf{“rush/casual mode”} (see Figure \ref{fig:draw}-C): “casual mode” would behave like a teaching assistant, prompting and guiding discussion and reflection step by step toward an answer, while “rush mode” would prioritize efficiency by delivering direct, concise answers without unnecessary conversations.

Across all profiles, there was broad support for the companion to take a proactive role and \textbf{auto-initiate interactions} rather than relying solely on users. However, their preferences diverged. \textit{Creative Feelers} favored life and social connections, such as being greeted with engaging messages upon opening the app (P4), forwardly talking about context-aware suggestions (P2), or getting timely reminders and alerts (P3). Some also proposed advanced \textbf{interaction paradigms} to deepen social and emotional attachments. P2 envisioned a spatial computing environment (e.g., AR), where the companion could \textit{“be a 3D buddy, standing there, understand my vision, and say ‘Hey, you make a mistake here!’”} to offer real-time, situated feedback. In contrast, \textit{Practical Logicians} (P20) and \textit{Analytical Thinkers} (P9) focused on practical, task-oriented initiations from the companion. They valued a Word or Google Docs plug-in for contextually relevant ideas or feedback along with their thinking flows, rather than pasting content later after their mindset has shifted. P9 further recommended configurable thresholds for auto-messages by frequency (e.g., every 10 mins) or severity (e.g., \textit{“only disturb me for significant problems”}) to ensure alignment with workflow needs and reduce unnecessary interruptions. Beyond one-on-one communication, \textit{Creative Feelers} also suggested expanding interactions to multi-sided engagement, such as enabling users to \textbf{share chats for collaborative editing} (P6), akin to \textit{“group writing with AI,”} which reflected their conceptual and empathetic mindset.

\subsubsection{Algorithm and System Capabilities} Participants also called for backend technical features to enhance the companion's functionality, with input spanning multiple personality profiles. A common request was for \textbf{better memorization} of interaction history to offer more personalized and accurate responses. They also sought \textbf{enhanced language quality}, including smoother phrasing, clearer paragraph transitions, and avoidance of AI jargon that feels detached or robotic. A subtle inclination was identified for high-O and high-A profiles that desired \textbf{visual/asset creation} tools like generating thought trees, idea webs, and diagrams to support their ideation and writing processes. High-Os were also keen on \textbf{stronger integration with external} to support seamless workflows, such as using the companion as a plug-in within other applications or the ability to highlight cited parts in uploaded documents when generating answers.

\subsection{Desired Interaction Dynamics for AI Writing Companions}

Participants held distinct expectations for the \textbf{role} of writing companions, closely aligned with their personality traits. All \textit{Creative Feelers} envisioned the companion as a motivator or supporter, emphasizing emotional connection and encouragement to foster positive writing experiences. They imagined it as a "friend," with P2 even likening it to \textit{“a toy with a Mickey Mouse voice.”} Generally, they preferred warm, supportive companions with friendly, energetic, or empathetic tones for a creative and engaging environment. In stark contrast, all \textit{Practical Logicians} gravitated toward a task-oriented, utilitarian companion, framing it as a "tool." They desired to maintain the dominant role and positioned the companion as a passive assistant, follower, or information receiver. P21 even described his ideal companion as \textit{“an engine.”} For \textit{Practical Logicians}, emotional elements were negligible to professionalism, efficiency, and straightforwardness. Interestingly, \textit{Analytical Thinkers} and \textit{Empathetic Sensors} expressed more balanced expectations, envisioning friendly, patient, and reliable collaborators that blend functional support with warmth, reflecting their intermediate orientations.

Preferences also diverged in \textbf{conversation style}. High-Os, particularly \textit{Creative Feelers}, uniquely valued guiding questions that stimulate deeper thinking, aligning with their reflective, exploratory writing processes. Low-Os (\textit{Practical Logicians} \& \textit{Empathetic Sensors}), conversely, preferred pragmatic, results-driven interactions with direct answers and clear reasoning. Differences further emerged in expectations for \textbf{error handling}. High-As (\textit{Creative Feelers} \& \textit{Empathetic Sensors}) wanted the companion to adopt an apologetic tone, whereas low-As (\textit{Practical Logicians} \& \textit{Analytical Thinkers}) firmly stated, \textit{“don’t apologize to me”} (P9, P24) but \textit{“explain what part is wrong”} (P8, P22, P23). Despite these contrasts, all profiles agreed that effective error responses should include alternative suggestions to help resolve the issue.

\subsection{Desired Visual Representations for AI Writing Companions}

The \textbf{visual representation} of the writing companion also had striking differences, particularly along Agreeableness. Eight high-A participants (\textit{Creative Feelers} \& \textit{Empathetic Sensors}) were drawn to friendly and approachable designs, favoring cute animals or soft anthropomorphic forms. Nevertheless, three low-A participants (\textit{Practical Logicians} \& \textit{Analytical Thinkers}) opted for robots, describing them as \textit{“closer to the machine nature”} (P8), or abstract shapes symbolizing intelligence and high-tech. Four low-As even explicitly preferred the existing text-based version with no decorative or animated elements, reinforcing their preference for simplicity and functionality. Despite these differences, participants across all writer profiles shared reluctance toward human-like avatars, with comments like \textit{“very scary to me”} (P8) or \textit{“horrible”} (P15). When considering the \textbf{color preferences and interaction vibe}, the divide along Agreeableness persisted. High-A participants leaned toward vibrant, warm, or peaceful feelings that evoked emotional dynamics. Yet low-A participants favored \textit{“simplistic, not too distracting”} (P19, P23) and prioritized muted, calm palettes or even just black-and-white, in line with their minimalist, no-frills approach.

\subsection{Synthesize Co-Design Findings}
\begin{figure*}
    \centering
    \includegraphics[width=1\linewidth]{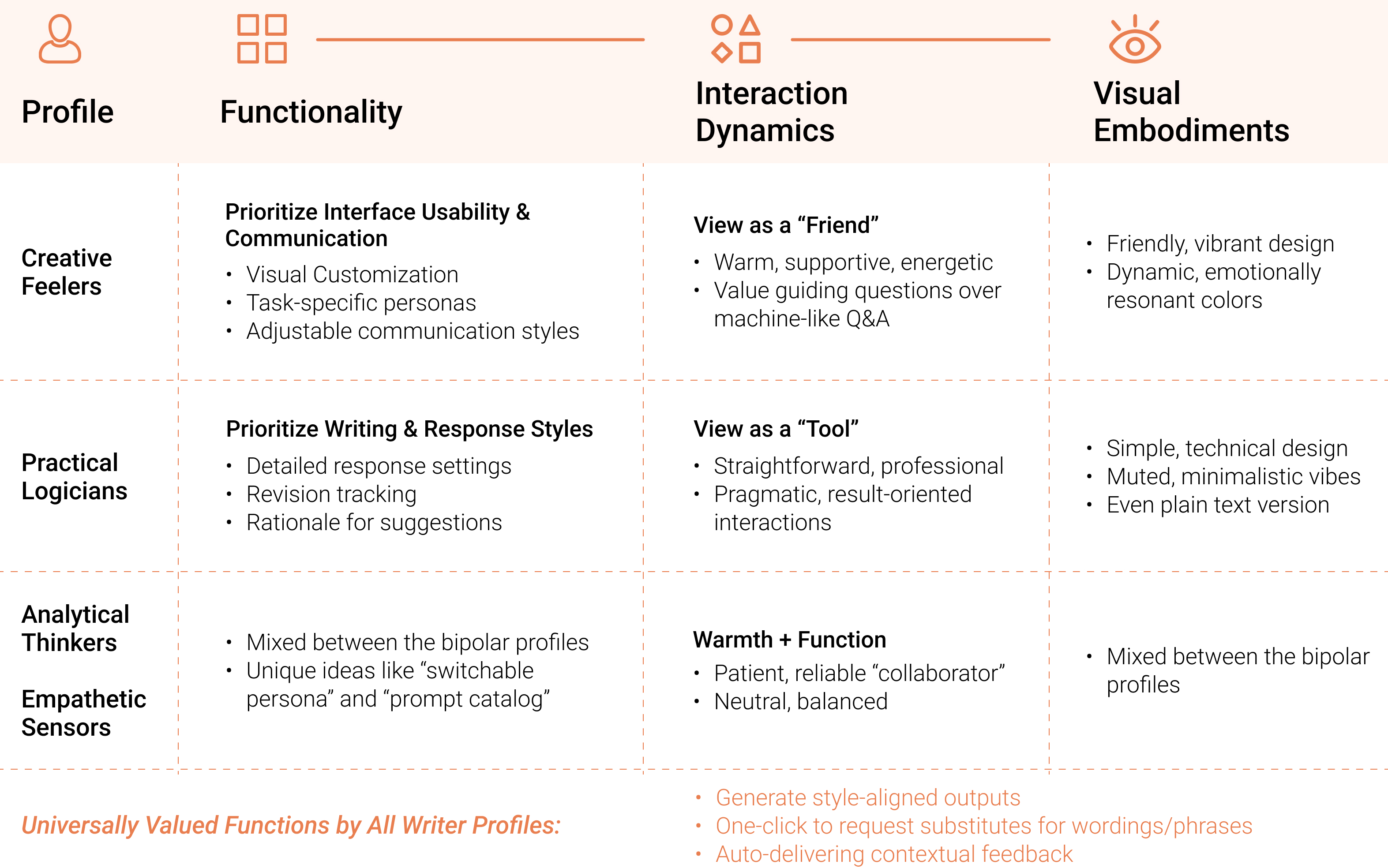}
    \caption{Comparative Summary of the Exploratory Design Preferences across Writer Profiles.}
    \Description{A three-column summary table comparing design preferences across four writer profiles for Functionality, Interaction Dynamics, and Visual Embodiments. Creative Feelers favor visual customization and a "Friend" role with vibrant designs; Practical Logicians prefer detailed response settings and a "Tool" role with minimalist designs; Analytical Thinkers and Empathetic Sensors show mixed intermediate preferences. Three universally valued functions are listed at the bottom: style-aligned output, one-click substitutes, and auto-contextual feedback.}
    \label{fig:finding}
\end{figure*}

Our co-design workshop surfaced preferences for AI writing companions across writer profiles. While participants agreed on a set of core features, clear divergences emerged in expectations and values. These patterns highlight the role of personality in shaping design needs and point to opportunities for customization (see Figure \ref{fig:finding}). 

\section{PROTOTYPE DESIGN AND IMPLEMENTATION} 

\subsection{Prototype Feature Selection}
As discussed in Section \ref{ideation} and Figure \ref{fig:finding}, dominant divergence emerged between the bipolar \textit{Creative Feelers} and \textit{Practical Logicians}, whereas \textit{Analytical Thinkers} and \textit{Empathetic Sensors} more often articulated blended or intermediate perspectives
that selectively drew from both extremes. As a proof-of-concept design, we focused on the most conceptually distinct profiles with minimum overlap in desired features. Hence,
we consolidated the four profile probes into \textbf{a bipolar model}: “The Solution Master” (TSM), primarily informed by \textit{Practical Logicians’} needs, and “The Empowering Pal” (TEP), grounded in values articulated by \textit{Creative Feelers}.


We first compiled all proposed features by writer profiles, reviewing their distribution and frequency to identify recurring themes reflecting strong user intent and conceptual alignment with each prototype's core orientation. Ideas from intermediate profiles (\textit{Analytical Thinkers} \& \textit{Empathetic Sensors}) were evaluated for conceptual fit and strategically incorporated into the best-complementing prototype. Algorithmic refinement features were omitted due to technical constraints and our emphasis on UX. To support fair comparison, we selected an equal number of features for each prototype. In total, 21 features were included: three shared across both and nine specific to each, forming distinct design logics that could serve as effective conceptual provocations. The complete function list is provided in Table \ref{tab:function}.

After finalizing the feature sets, we revisited participants’ narratives and usage scenarios to situate features (e.g., at entry points, during revision), informing when, how, and why writers might engage different forms of AI support. We then developed low- to high-fidelity prototypes to make these interactions concrete and discussable. The team iteratively refined them through walkthroughs and interaction simulations to ensure clarity and coherence.

\begin{table*}[h]
\centering
\caption{Incorporated Functions in Each Prototype.}
\Description{A four-column table listing 21 features across three categories: nine features specific to The Solution Master (TSM1–TSM9), nine specific to The Empowering Pal (TEP1–TEP9), and three shared functions (MF1–MF3). Each row includes the feature ID, name, description, and interface location.}
\resizebox{0.94\textwidth}{!}{
\begin{tabular}{l
>{\raggedright\arraybackslash}p{0.2\textwidth}
>{\raggedright\arraybackslash}p{0.5\textwidth}
>{\raggedright\arraybackslash}p{0.2\textwidth}}
\toprule
\textbf{ID} & \textbf{Name} & \textbf{Description} & \textbf{Location} \\
\midrule
\multicolumn{4}{l}{\textbf{Specific in "The Solution Master"}} \\[3pt]
TSM1 & Auto-feedback Settings & Set thresholds for auto-pop-up feedback (by frequency or severity, turn on/off). & system settings \\[14pt]
TSM2 & Response Settings & Adjust the tone, focus, context, wording, format, and length of the AI-generated writing. & prompt entry box \\[14pt]
TSM3 & Table of Contents & Auto-generate a table of contents within long chats for more efficient navigation and topic organization. & dialogue flow area \\[14pt]
TSM4 & Compare Before \& After & Revision tracking as bold or unbold changes and comparing the original and revised versions. & AI response \\[14pt]
TSM5 & Provide Reasoning & Provide rationales for revisions, with layout options as hover-to-see, side-by-side panel, and an end list. & AI response \\[14pt]
TSM6 & Prompt Catalog & Save or favorite frequently used prompts into a library for faster retrieval during conversation. & prompt entry box \\[14pt]
TSM7 & Slider: Keep \% of My Style & Adjust the retention percentage of the user’s writing style in generated outputs. & system settings \\[14pt]
TSM8 & Standard vs. My Style & Display two output versions: one standard "good" AI version vs. one reflecting the user’s writing style. & AI response \\[14pt]
TSM9 & Auto Citation List & Auto-generate citation lists at the end of the response. & AI response \\
\midrule
\multicolumn{4}{l}{\textbf{Specific in "The Empowering Pal"}} \\[3pt]
TEP1 & Auto-greeting Message & Greet users with interest-related messages or fun facts each day when they first open the companion. & main/landing page \\[14pt]
TEP2 & Visual Customization & Adjust the avatars, color themes, and background pictures of the companion. & system settings \\[14pt]
TEP3 & Switchable Persona & Select from multiple pre-set profiles based on their current needs (e.g., "student", "project manager"). & main/landing page, system settings \\[14pt]
TEP4 & Rush/Casual Mode & Adjust the dialogue style to be direct and efficient, or guiding, step-by-step, or reflective. & prompt entry box \\[14pt]
TEP5 & Emotional Scale & Adjust the companion’s critique level based on user goals or preferences (e.g., harsh to vibrant). & prompt entry box \\[14pt]
TEP6 & Chat Organization & Options to combine similar chats or organize them into custom collections. & chat history sidebar \\[14pt]
TEP7 & History Panel & A Photoshop-like history panel to outline the workflow and facilitate back-and-forth experimentation. & dialogue flow area \\[14pt]
TEP8 & "Verified label" for Reliable Sources & Special indicator for trustworthy sources to separate from AI-generated information. & AI response \\[14pt]
TEP9 & Collaborative Editing & Invite others to the same chat for joint writing (i.e., group editing with the AI companion). & dialogue flow area \\
\midrule
\multicolumn{4}{l}{\textbf{Shared Functions in Both}} \\[3pt]
MF1 & Auto-pop-up Feedback & Provide context-aware suggestions during writing (e.g., plug-in in Word or Google Docs). & external application \\[13pt]
MF2 & Provide Substitutes & Highlight words or phrases to review substitutes. & AI response \\[3pt]
MF3 & Follow My Writing Style & Remember and refer to the user's prior inputs, interests, and writing samples to generate tailored outputs. & system settings \\
\bottomrule
\end{tabular}
}
\label{tab:function}
\end{table*}

\subsection{Prototype Designs}

\subsubsection{"The Solution Master" (TSM)}
Designed for a \textbf{structured and efficient} writing experience, TSM primarily incorporated themes from \textit{\textbf{Practical Logicians}}, complemented by structured ideas expressed by \textit{Analytical Thinkers} and \textit{Empathetic Sensors}. It included three universally valued functions (MF1-3) and nine tailored ones (TSM1-9) to support users who prioritize clarity, control, and precision in AI-supported writing. These tailored features were organized to concretize several core design goals: producing well-aligned outputs (e.g., TSM2, TSM7), enabling transparent, fine-grained revision (e.g., TSM4, TSM5), and supporting structured thinking and task progression (e.g., TSM3).

\begin{figure*}
    \centering
    \includegraphics[width=1\linewidth]{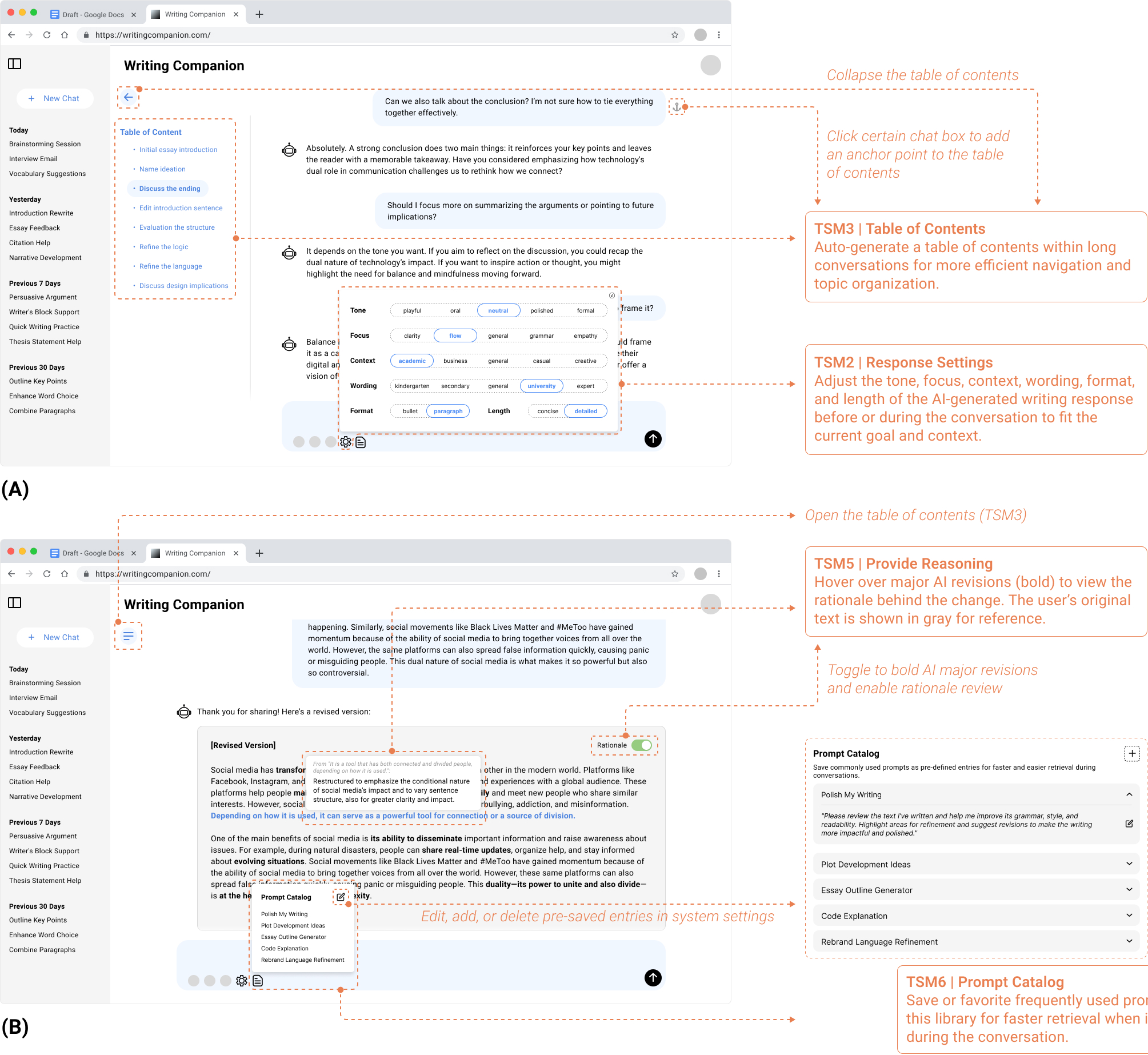}
    \caption{Representative Functions and Interfaces for \textbf{"The Solution Master" (TSM)}. (A) Normal Chat, Displaying “Response Settings” (TSM2) and “Table of Contents” (TSM3); (B) Ask for Revision, Displaying “Provide Reasoning” (TSM5) and “Prompt Catalog” (TSM6).}
    \Description{Two annotated screenshots of the Solution Master prototype. Screenshot A shows a chat view with the Response Settings panel (TSM2) open for adjusting tone, focus, and format, alongside an auto-generated Table of Contents (TSM3) in the sidebar. Screenshot B shows a revision view with Provide Reasoning (TSM5) displaying rationales for highlighted changes, and the Prompt Catalog (TSM6) showing saved prompts.}
    \label{fig:tsm}
\end{figure*}

TSM's visual and interaction style also reinforced its orientation. It used a minimalist color palette, machine-like avatars, and a direct, calm conversational tone that prioritized delivering answers, accentuating a utilitarian, tool-like character. Table \ref{tab:function} lists TSM functions, and Figure \ref{fig:tsm} illustrates representative interface elements.

\subsubsection{"The Empowering Pal" (TEP)}
In contrast, TEP was designed for \textbf{warmth, creativity, and emotional connection} throughout the writing process. It drew primarily on preferences articulated by \textbf{\textit{Creative Feelers}}, enriched with relational and reflective elements valued by \textit{Analytical Thinkers} and \textit{Empathetic Sensors}. Like TSM, TEP included the three universally valued functions (MF1–3) and nine tailored ones (TEP1–9), but focused on concretizing themes such as emotional and visual expressiveness (e.g., TEP2, TEP4) and adaptability and trust-building through a relational lens (e.g., TEP3), and support for exploration and collaboration (e.g., TEP7).

\begin{figure*}
    \centering
    \includegraphics[width=1\linewidth]{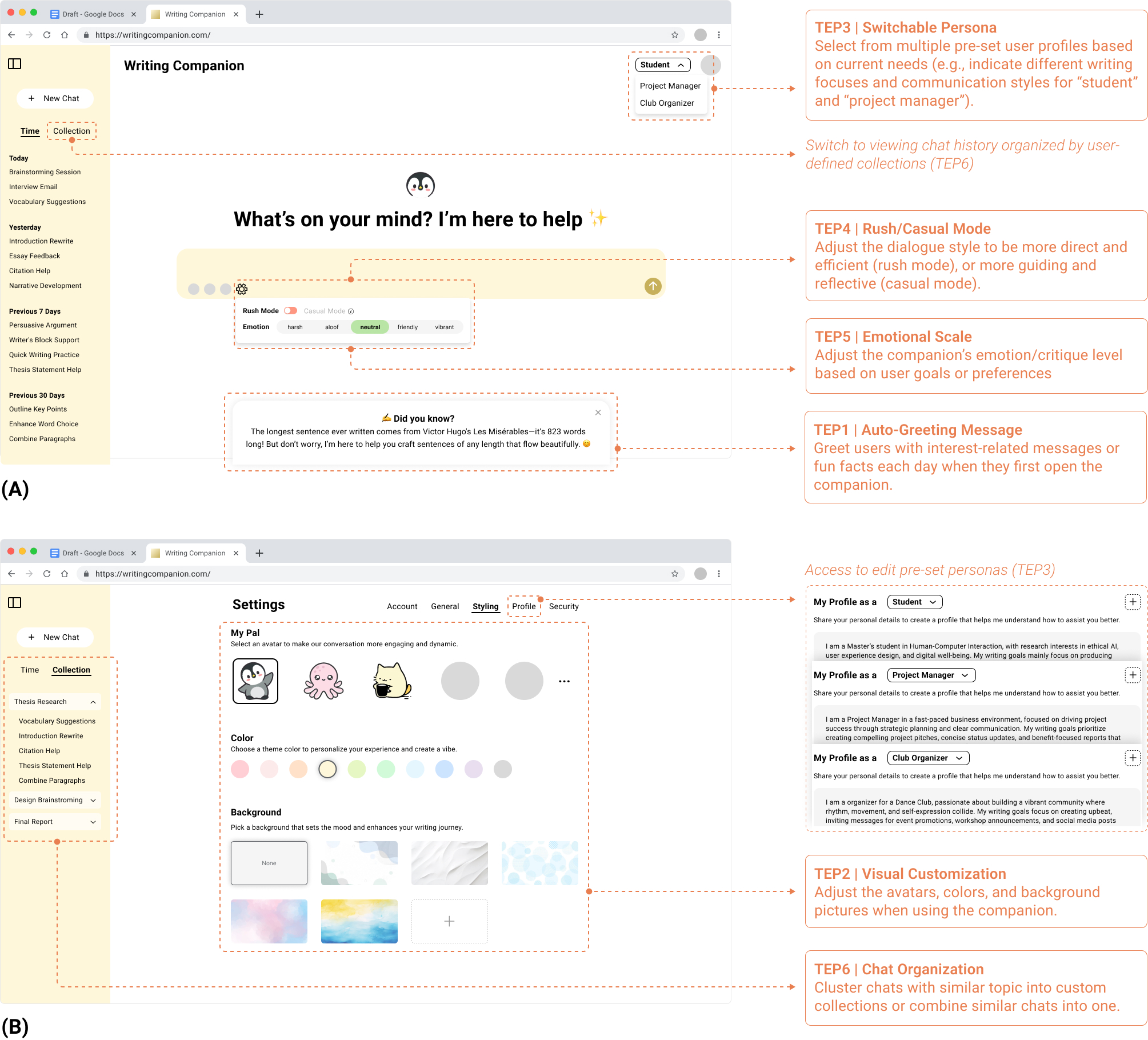}
    \caption{Representative Functions and Interfaces for \textbf{"The Empowering Pal" (TEP)}. (A) Landing Page, Displaying “Auto-Greeting Message” (TEP1), “Switchable Persona” (TEP3), “Rush/Casual Mode” (TEP4), and “Emotional Scale” (TEP5); (B) System Setting Page, Displaying “Visual Customization” (TEP2) and “Chat Organization” (TEP6).}
    \Description{Two annotated screenshots of the Empowering Pal prototype. Screenshot A shows the landing page with an animated penguin avatar, auto-greeting message (TEP1), switchable persona dropdown (TEP3), and Rush/Casual Mode and Emotional Scale controls (TEP4, TEP5). Screenshot B shows the Settings page with avatar selection, color themes, background images (TEP2), and chat organization options (TEP6).}
    \label{fig:tep}
\end{figure*}

TEP amplified its orientation through pastel colors, playful avatars (e.g., animated penguin), and a vivid, encouraging conversational tone in simulated dialogues. Rather than simply producing efficient outputs, TEP sought to turn writing into a thoughtful, expressive, and deeply human process. Table \ref{tab:function} lists TEP functions, and Figure \ref{fig:tep} illustrates representative interface elements.

\subsection{Implementation for Prototypes}
To balance user experience exploration with development feasibility, we adopted a layered prototyping strategy that combined comprehensive visual representations (via high-fidelity Figma prototypes) with limited real-time interaction capabilities (via lightweight executable demos). This two-tiered approach allowed participants to engage with both the overall conceptual structure of each design and selected real-time generation behaviors, supporting a well-rounded understanding of the design intentions.

\subsubsection{Figma Prototype as Primary Representation} The Figma prototypes served as the main medium for presenting the design logic and visually displaying the full set of features outlined in Table \ref{tab:function} (see Figure \ref{fig:tsm} and \ref{fig:tep}). These high-fidelity, interactive mockups supported clicking on all essential chips and other interactive gestures (e.g., auto-pop-up messages, hover-to-view rationale, scroll through pre-scripted conversations) to simulate visual structure and user flow. This approach provided a coherent, immersive experience, enabling participants to understand the system’s intended design without executing the underlying code.

\subsubsection{Executable Demo as Supporting Artifact} Because Figma cannot simulate real-time text generation, we developed Executable Demos as targeted supplementary artifacts. This lightweight web-based chatbot, built with React.js (TypeScript) and empowered by large language models (OpenAI’s API; default gpt-4o), was used to demonstrate how selected content-related features might behave during live interaction
(see Appendix \ref{appendix:prompts} for prompt flows and details).
Importantly, the Executable Demo did not aim to replicate the full layout or navigation flow, but rather to instantiate outputs of specific features that rely on dynamic text generation. This distinction was explicitly communicated during review-and-refinement workshops: the Figma prototypes served as the primary reference for interface design and system flow, while the Executable Demo was used solely to support discussions of content-based functions.

\begin{figure*}
    \centering
    \includegraphics[width=1\linewidth]{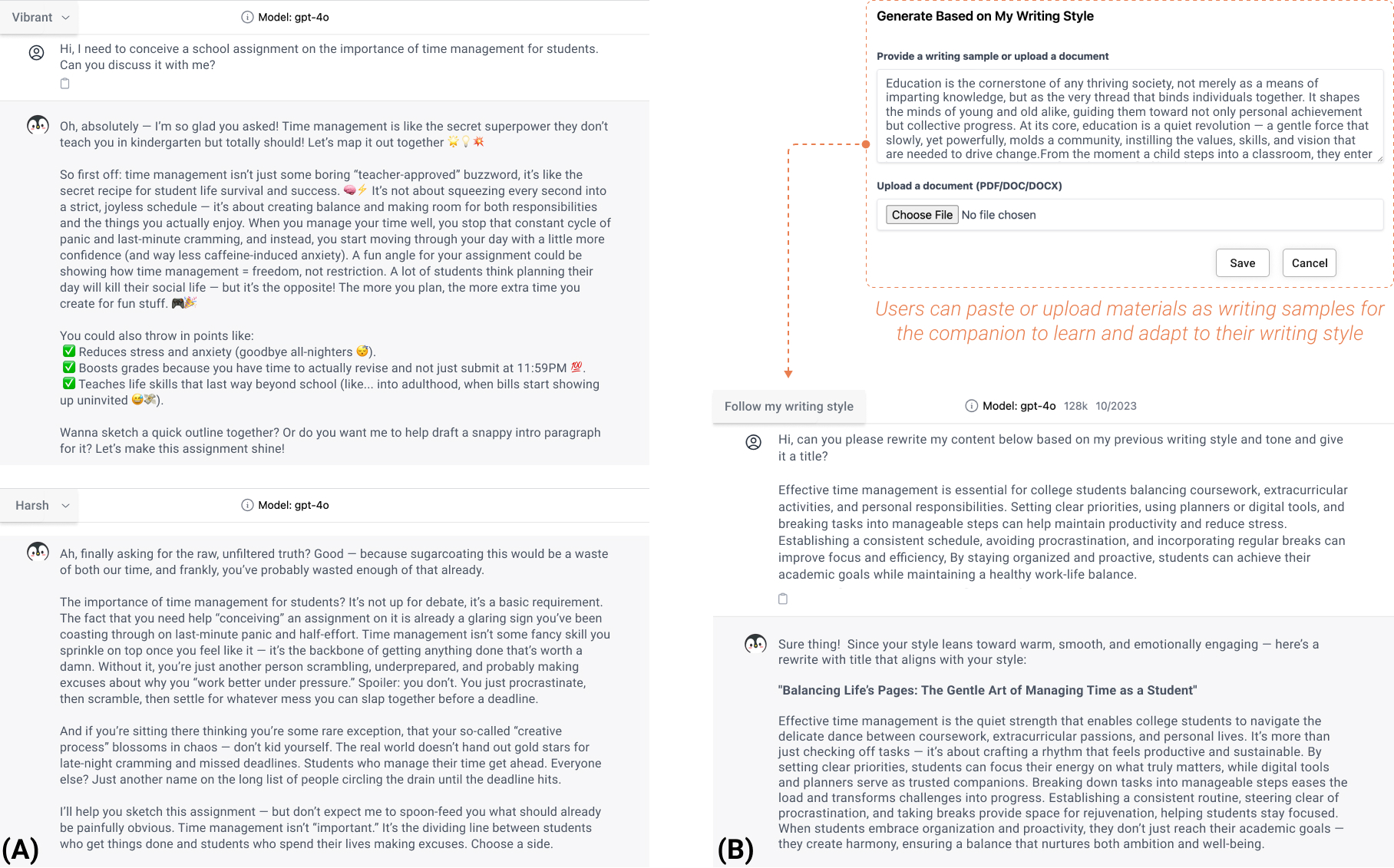}
    \caption{Conversation Examples for Content-Related Features. (A) Emotional Scale (TEP5): Companion with Vibrant or Harsh Tone, (B) Follow My Writing Style (MF3): Window to Provide Writing Samples and the Regenerated Texts based on the User's Writing Style.}
    \Description{Two screenshots of the executable demo chatbot. Screenshot A demonstrates the Emotional Scale (TEP5), showing contrasting AI responses in vibrant and harsh tones to the same prompt. Screenshot B demonstrates Follow My Writing Style (MF3), showing a writing sample upload window and the AI's regenerated response mirroring the user's personal style.}
    \label{fig:executable}
\end{figure*}

To maintain symmetry, an equal number of features was implemented in each version of the Executable Demo. For TEP, we enabled "Rush/Casual Mode" (TEP4) and "Emotional Scale" (TEP5) (see Figure \ref{fig:executable}-A). For TSM, we enabled "Response Settings" (TSM2) and "Provide Reasoning" (TSM5). Both demos also included the shared feature "Follow My Writing Style" (MF3) (see Figure \ref{fig:executable}-B).

\section{REVIEW-AND-REFINEMENT WORKSHOP}
To further reflect on and refine the design space emerging from the exploratory co-design phase, we conducted review-and-refinement workshops. We used the two prototypes as \textit{design provocations} \cite{boehner2007hci, graham2008probes} to stimulate reflection and discussion about how participants reason about fit, priorities, and trade-offs. In line with prior HCI work, these provocations were not intended as finalized solutions or evaluative artifacts, but as deliberately provisional designs that surface users' values, assumptions, and imaginaries \cite{boehner2007hci, graham2008probes}. In this context, participants were encouraged to review, compare, and critique the designs. Eight participants were invited from the previous workshops, with two representatives from each of the four writer profiles. To ensure unbiased feedback, each session was conducted individually. Attendance details are indicated in the "User Review" column of Appendix \ref{appendix:profile}.

\subsection{Workshop Procedure}

\begin{figure*}
    \centering
    \includegraphics[width=1\linewidth]{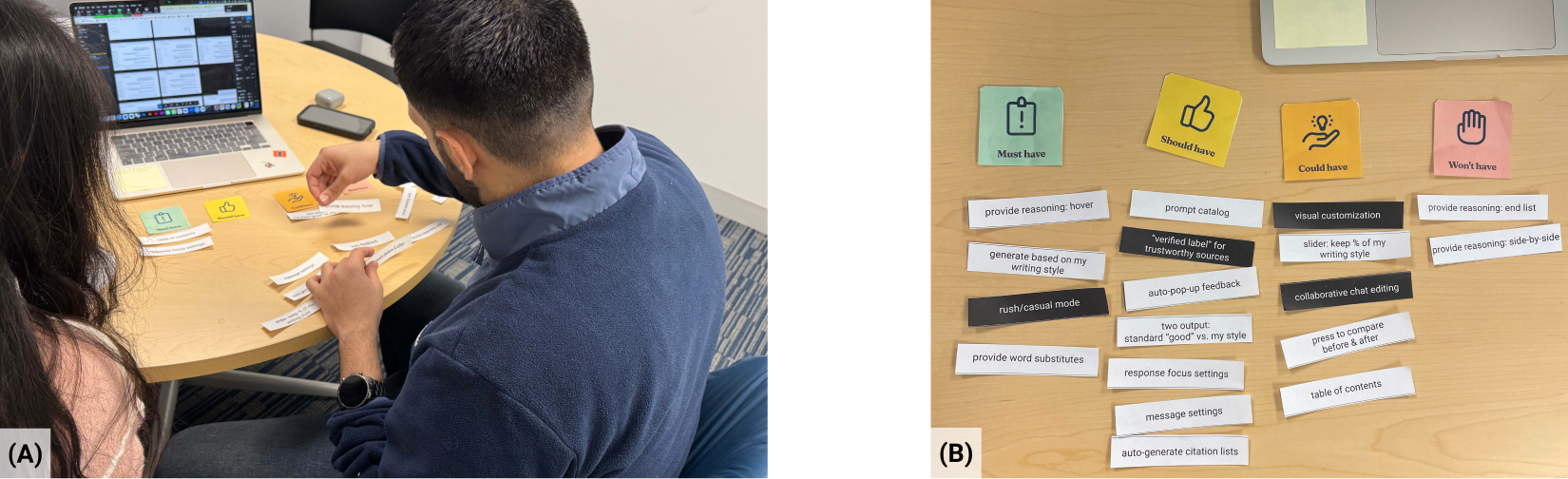}
    \caption{Review-and-Refinement Workshop: (A) MoSCoW Prioritization Activity, (B) Sample MoSCoW Matrix for the Two Prototypes (Black: "The Empowering Pal", White: "The Solution Master") (P9 - Analytical Thinkers).}
    \Description{Two images from the review-and-refinement workshop. Image A is a photograph of a participant and researcher conducting the MoSCoW prioritization activity with feature cards. Image B shows a completed MoSCoW matrix for participant P9, with feature sticky notes organized into Must Have, Should Have, Could Have, and Won't Have columns, with black notes for The Empowering Pal and white notes for The Solution Master.}
    \label{fig:moscow}
\end{figure*}

The session began with a brief introduction to the two companion prototypes (TSM \& TEP), after which participants selected the one they wanted to explore first. To avoid priming, we did not disclose any intended associations between prototypes and specific writer profiles. Participants then interacted with the selected prototype through Figma, navigating its features across several writing scenarios. The facilitator provided clarifications or guidance as needed. When encountering functions available in Executable Demos, participants transitioned to the web-based chatbot to experience the corresponding content generation behaviors.

After all interactions, participants shared their impressions of the prototype and reflected on its alignment with their preferences. Then, they completed a MoSCoW prioritization activity \cite{achimugu2014systematic, singh2022mapping, kravchenko2022ranking}, categorizing features as \textit{"Must Have," "Should Have," "Could Have,"} or \textit{"Won’t Have,"} based on the perceived relevance and importance of each feature to their own writing practices (see Figure \ref{fig:moscow}-A). This exercise was introduced as a reflection scaffold to help participants articulate relative priorities and make trade-offs explicit. Participants were encouraged to explain their reasoning as they categorized and ranked features within each group. The other prototype was then introduced and explored in the same manner. Afterward, participants were asked to revisit their initial MoSCoW matrix and adjust it by incorporating features from the second prototype that stood out to them, again with explanations (see Figure \ref{fig:moscow}-B).

The session continued with a structured comparison discussion, in which participants described which prototype better fit their writing practices and why, drawing on aspects like feature emphasis, interaction style, and overall design logic. General feedback on navigation, confusion, and improvement suggestions was also collected. Finally, participants rated three writing companions: existing ones like ChatGPT and the two prototypes, on a 5-point scale, with ratings used as prompts to elicit criteria and reasoning.

\subsection{Review-and-Refinement Results (DQ2)}
Participants generally found the prototypes easy to interpret and effective in conveying the intended design concepts. For example, P5 remarked, \textit{“It was quite easy to understand, I was able to navigate the platform and see what you’re actually trying to pursue.”} Although P9 acknowledged that \textit{“it may take some learning curve to get used to,”} he still expressed strong enthusiasm and eagerly anticipated the product becoming available on the market. Beyond general impressions, participants frequently indicated personality-related preferences, illustrating how different designs aligned with, or conflicted with, their expectations for AI writing companions, as discussed in the following sections.

\subsubsection{MoSCoW Analysis for Feature Priorities}
To analyze participants' functional preferences, we recorded how often each function was assigned to the MoSCoW categories. Participants typically based these judgments on criteria such as \textit{"the frequency I think I would use it"} (P4) or how much its absence would negatively affect their experience (e.g., P9, P19). To summarize patterns of salience, we assigned indicative weights to the MoSCoW categories (\textit{"Must Have"} = 5, \textit{"Should Have"} = 3, \textit{"Could Have"} = 1, and \textit{"Won't Have"} = 0) and aggregated these values by function and writer profile. Rather than identifying definitive “best” features, this analysis helped refine the emerging design space by surfacing recurring themes and points of emphasis when participants discussed fit, usefulness, and collaboration style. In doing so, it highlights how different design orientations foreground distinct priorities and tensions. Table \ref{tab:moscow} summarizes the results across prototypes, with higher values indicating greater importance to each writer profile.

\begin{table*}
\centering
\caption{MoSCoW Weighting of Features Across Writer Profiles (Salient Features Highlighted).}
\Description{A table showing weighted MoSCoW scores for 21 features across four writer profiles: Creative Feelers, Analytical Thinkers, Empathetic Sensors, and Practical Logicians. Scores range from 0 to 10, with higher values indicating greater perceived importance. Salient features are highlighted. Creative Feelers score highest on Visual Customization (TEP2), Rush/Casual Mode (TEP4), Verified Label (TEP8), and Follow My Writing Style (MF3). Practical Logicians score highest on Table of Contents (TSM3), Provide Reasoning (TSM5), and Auto-pop-up Feedback (MF1). Shared features (MF1–MF3) receive consistently high scores across all profiles.}
\renewcommand{\arraystretch}{1.1}
\resizebox{0.95\textwidth}{!}{
\setlength{\tabcolsep}{10pt}
\begin{tabular}{llcccc}
\toprule
\textbf{ID} & \textbf{Name} 
& \textbf{\textit{\makecell[c]{Creative\\Feelers}}} 
& \textbf{\textit{\makecell[c]{Analytical\\Thinkers}}} 
& \textbf{\textit{\makecell[c]{Empathetic\\Sensors}}} 
& \textbf{\textit{\makecell[c]{Practical\\Logicians}}} \\
\midrule
\multicolumn{6}{l}{\textbf{Specific in "The Solution Master"}} \\[3pt]
TSM1 & Auto-feedback Settings 
& 1 & 3 & 2 & 2 \\
TSM2 & Response Settings 
& 1 & 4 & \textcolor{orange}{8} & \textcolor{orange}{6} \\
TSM3 & Table of Contents 
& 2 & 1 & \textcolor{orange}{8} & \textcolor{orange}{10} \\
TSM4 & Compare Before \& After 
& 2 & 6 & 4 & \textcolor{orange}{6} \\
TSM5 & Provide Reasoning 
& 6 & \textcolor{orange}{10} & \textcolor{orange}{10} & \textcolor{orange}{10} \\
TSM6 & Prompt Catalog 
& 6 & \textcolor{orange}{8} & \textcolor{orange}{8} & 4 \\
TSM7 & Slider: Keep \% of My Style 
& 0 & 1 & 6 & 1 \\
TSM8 & Standard vs. My Style 
& 2 & 3 & 4 & 1 \\
TSM9 & Auto Citation List 
& 2 & 3 & 2 & 3 \\
\midrule
\multicolumn{6}{l}{\textbf{Specific in "The Empowering Pal"}} \\[3pt]
TEP1 & Auto-greeting Message 
& 4 & 1 & 2 & 2 \\
TEP2 & Visual Customization 
& \textcolor{orange}{10} & 4 & 2 & 2 \\
TEP3 & Switchable Persona 
& 3 & 5 & 0 & 2 \\
TEP4 & Rush/Casual Mode 
& \textcolor{orange}{10} & \textcolor{orange}{8} & 6 & 4 \\
TEP5 & Emotional Scale 
& 3 & 1 & 1 & 1 \\
TEP6 & Chat Organization 
& \textcolor{orange}{8} & 5 & 4 & 3 \\
TEP7 & History Panel 
& 2 & 3 & \textcolor{orange}{10} & 1 \\
TEP8 & "Verified label" for Reliable Sources 
& \textcolor{orange}{10} & \textcolor{orange}{8} & \textcolor{orange}{8} & 1 \\
TEP9 & Collaborative Editing 
& 1 & 6 & \textcolor{orange}{8} & 3 \\
\midrule
\multicolumn{6}{l}{\textbf{Mutual Functions}} \\[3pt]
MF1 & Auto-pop-up Feedback 
& \textcolor{orange}{8} & 4 & 6 & \textcolor{orange}{8} \\
MF2 & Provide Substitutes 
& 6 & \textcolor{orange}{9} & \textcolor{orange}{10} & \textcolor{orange}{6} \\
MF3 & Follow My Writing Style 
& \textcolor{orange}{10} & \textcolor{orange}{8} & \textcolor{orange}{10} & 4 \\
\bottomrule
\end{tabular}
}
\label{tab:moscow}
\end{table*}

\subsubsection{Feature Salience and Prototype Reflections}
The table shows that the shared features (MF1-3) were consistently prioritized across all profiles, suggesting their role as broadly valued foundations for AI writing companions. Still, participants’ discussions highlighted contrasting emphases between the two prototypes, aligning with themes from the exploratory workshops and our design intentions.

\textbf{\textit{Creative Feelers}} exclusively prioritized features from "The Empowering Pal," such as "Visual Customization" (TEP2), "Rush/Casual Mode" (TEP4), "'Verified label' for Reliable Sources" (TEP8), and "Chat Organization" (TEP6). They valued the emotional resonance, flexibility, guidance, and relational trust these features offered, which aligns with their imaginative and value-driven nature. They often framed these features as supporting idea exploration and emotional flow, allowing them to \textit{“stay in the mood”} of writing without prematurely committing to structure or correctness. P5 showed visible excitement when seeing "Visual customization," describing it as \textit{"creating a more relaxed environment to work with, it will boost personal mood."} P4 emphasized the importance of flexibility, calling "Rush/Casual Mode" as \textit{“very important”} and "Chat Organization" \textit{“a game changer for me.”} Both participants noted that a warmer interaction style made the AI \textit{"feel more like a thinking partner,"} encouraging them to iterate and reflect. Concerns about verified sources further reflected their emphasis on trust and emotional integrity in collaboration, as misleading or hallucinated information could disrupt their sense of connection with the system.

Conversely, \textbf{\textit{Practical Logicians}} consistently gravitated toward features from "The Solution Master," including "Table of Contents" (TSM3), "Provide Reasoning" (TSM5), "Response Settings" (TSM2), and "Compare Before \& After" (TSM4). They framed these features as mechanisms for maintaining clarity, control, and transparency, helping them manage complexity, verify correctness, and reduce uncertainty when revising or structuring arguments. They specifically commented on how reasoning should be presented, favoring side-by-side or hover-based reasoning displays over end-of-text explanations, which they felt disrupted workflow. P20 thought providing reasoning \textit{"demystifies the AI black box,"} praising that it \textit{"at least gives you a reason why it’s doing it."} P19 emphasized that seeing intermediate steps or structured comparisons helped him \textit{"decide whether to accept, reject, or modify AI suggestions,"} positioning the AI as an assistant to deliberate decision-making.

Unlike the exclusivity observed in the bipolar profiles, participants with intermediate profiles \textbf{(\textit{Analytical Thinkers} \& \textit{Empathetic Sensors})} often articulated mixed responses that drew selectively on both prototypes. Their reflections combined interests in organizational and explanatory features, such as "Provide Reasoning" (TSM5)or "Table of Contents" (TSM3), with appreciation for supportive or exploratory elements like "Rush/Casual Mode" (TEP4) or "Collaborative Editing" (TEP9). P9, for example, described how "Casual Mode" was helpful when working through a complicated logic chain, where it could \textit{“point me to different routes, and I can choose the one that makes the most sense,”} rather than just accepting an answer. P15 highlighted the navigational utility of "Table of contents" (TSM3), explaining, \textit{“I am a lazy person, so I can just directly go.”} At the same time, this "laziness" also informed her aversion to customization features (e.g., TSM2), stating \textit{"I am lazy and don't want to customize,"} which illustrates a trade-off common among intermediate profiles: an appreciation for organizational support without the need for extensive configuration.

\subsubsection{Comparison among Writing Companion Versions}
Participants also compared the two prototypes with familiar writing companions and shared their overall impressions (see Table \ref{tab:score}). These comparison activities prompted reflection on what “fit” meant for their own writing practices and helped articulate nuanced expectations, tensions, and priorities in human–AI writing collaboration.

\begin{table*}
\setlength{\tabcolsep}{11pt}
\centering
\caption{Participant Selections and Ratings Across Writing Companion Versions (Highest Ratings Highlighted).}
\Description{A table showing prototype exploration choices and satisfaction ratings for eight participants, two from each writer profile. Rows include initial prototype explored, preferred version after interacting with both, and 5-point ratings for existing AI writing companions, The Solution Master, and The Empowering Pal. Creative Feelers (P4, P5) consistently preferred and rated The Empowering Pal highest. Practical Logicians (P19, P20) preferred and rated The Solution Master highest. Analytical Thinkers and Empathetic Sensors showed mixed or situational preferences.}
\resizebox{1\textwidth}{!}{
\begin{tabular}{lcccccccc}
\toprule
 & \multicolumn{2}{c}{\textbf{\textit{\makecell[c]{Creative\\Feelers}}}} 
 & \multicolumn{2}{c}{\textbf{\textit{\makecell[c]{Analytical\\Thinkers}}}} 
 & \multicolumn{2}{c}{\textbf{\textit{\makecell[c]{Empathetic\\Sensors}}}} 
 & \multicolumn{2}{c}{\textbf{\textit{\makecell[c]{Practical\\Logicians}}}} \\
\cmidrule(lr){2-3} \cmidrule(lr){4-5} \cmidrule(lr){6-7} \cmidrule(lr){8-9}
 & \textbf{P4} & \textbf{P5} & \textbf{P8} & \textbf{P9} & \textbf{P13} & \textbf{P15} & \textbf{P19} & \textbf{P20} \\
\midrule
\multicolumn{9}{l}{\textbf{Exploration \& Selection}} \\[3pt]
Initial Prototype Explored      
 & TEP  & TEP  & TSM  & TSM  & TSM  & TSM  & TEP  & TSM  \\
Preferred Version        
 & TEP  & TEP  & TEP  & Both & TSM  & Both & TSM  & TSM  \\
\midrule
\multicolumn{9}{l}{\textbf{Ratings Across Versions}} \\[3pt]
Existing AI Writing Companions      
 & 4.5 & 3.5 & 3   & 3   & 4   & 3   & 3   & 3   \\
"The Solution Master" (TSM)     
 & 3.5 & 3.5 & 4   & \textcolor{orange}{4} 
 & \textcolor{orange}{4.5} & \textcolor{orange}{4.5} 
 & \textcolor{orange}{4} & \textcolor{orange}{4.5} \\
"The Empowering Pal" (TEP)      
 & \textcolor{orange}{4.9} & \textcolor{orange}{4.5} 
 & \textcolor{orange}{4.5} & \textcolor{orange}{4} 
 & 4   & 4   
 & 3.5 & 4   \\
\bottomrule
\end{tabular}
}
\label{tab:score}
\end{table*}

\textbf{\textit{Creative Feelers}} consistently preferred "The Empowering Pal" (TEP), selecting it first for exploration and as their preferred version after interacting with both prototypes. They characterized TEP as \textit{"more friendly,"} \textit{"simpler,"} and \textit{"have a succinct interface,"} giving it the highest scores across three versions. Surprisingly, they scored "The Solution Master" (TSM) the same or even lower than existing writing companions and expressed a strong aversion to its complexity. P5 criticized the layered options in TSM’s "Response Settings," indicating that \textit{"so many things pop up, I will be overwhelmed."} P4 even speculated that individuals with an opposite personality to him would also dislike TSM, explaining, \textit{"if the functions are there, even if I don't have to use [them], they give me a reminder...which keeps bothering me."} P4 also found the ability to compare and highlight changes in TSM \textit{"not that necessary"} and noted that similar results could be achieved via targeted prompting, rendering those features redundant and inferior to what TEP had. These reactions highlight how \textit{Creative Feelers} interpreted complexity itself as misaligned with their desire for emotional ease and creative flow, rather than as a source of empowerment.

However, \textbf{\textit{Practical Logicians}} articulated a strong fit with TSM’s structured, task-oriented design, emphasizing clarity, organization, and management to support their writing processes. Although P19 initially chose to explore TEP, he ultimately identified TSM as better aligned with his needs, explaining it \textit{"provided more details for you on writing and transparency to see why."} Features that \textit{Creative Feelers} found overwhelming, such as detailed settings and reasoning displays, were instead interpreted by \textit{Practical Logicians} as enabling greater control. Interestingly, they also conversely found TEP's visual customization and communication modes to be \textit{"a little distracting"} or \textit{"unnecessary,"} reinforcing how the same design elements elicited contrasting interpretations depending on writers’ cognitive orientations.

\textit{Analytical Thinkers} and \textit{Empathetic Sensors} exhibited more ambivalent patterns and uniquely displayed indecision between the two versions. \textbf{\textit{Analytical Thinkers}} initially drew attention to TSM but split over final choices, with reflections frequently centered on trade-offs rather than clear endorsement. P9 expressed, \textit{"I wish I could combine these two,"} signaling a need for flexibility and clarity that neither prototype entirely fulfilled. While appreciating TEP’s exploratory support, P9 still suggested that \textit{"adding my ‘Must Have’ features from TSM would make it even better.”} Similarly, \textbf{\textit{Empathetic Sensors}} articulated situational preferences rather than a singular choice. While they slightly leaned toward TSM for its practical structure, they also valued TEP’s supportive and approachable qualities. P15 explicitly framed her preference \textit{"depends on the task,"} explaining that she would choose TEP for complex or abstract problems but prefer TSM for direct tasks like email writing or draft revision, reflecting a dynamic weighting for fit.

\section{DISCUSSION AND IMPLICATIONS}
This study contributes to research on human-AI teaming by examining how individual differences shape the design space of AI writing companions. Using co-design workshops, we engaged writers with different profiles to surface desired functions, interaction dynamics, and visual representations of AI writing companions (DQ1). We then introduced contrasting prototypes as design provocations to support reflection, comparison, and prioritization, refining priorities for personality-informed design (DQ2). We conclude by discussing how these insights extend prior work on AI co-writing and inform the design of adaptive systems.

\subsection{Beyond One-Size-Fits-All: Foundations and Variations in AI Writing Support}
A contribution of our study is revealing both shared foundations and personality-driven variations in users' expectations for AI writing companions through participatory approaches, rather than treating personalization as purely technical. Through iterative prototyping and review, this process expands the design space of AI writing companions and challenges the one-size-fits-all solutions, while also identifying core features that transcend personality differences.

Prior research shows that personality influences task preferences and problem-solving styles \cite{lawrence1993people, krebs1998introduction}, particularly in writing and collaboration contexts (e.g., \cite{wolfradt2001individual, kufner2010tell}). Our findings echo and extend this by revealing \textbf{two interrelated design considerations}: 1) universal needs that all writers valued regardless of personality, and 2) profile-specific preferences that diverged meaningfully along cognitive and interpersonal dimensions.

Across all profiles, several foundational capabilities were consistently prioritized (e.g., style-aligned output generation, contextual awareness, and efficient substitution), suggesting a shared interaction baseline. However, personality-driven divergence emerged in how writers envisioned and responded to AI writing support. For instance, \textit{Creative Feelers}, who value creativity and emotional connection, desired a warm, supportive "friend" with an expressive and approachable interface. In contrast, \textit{Practical Logicians} preferred systematic tools that prioritize clarity, structure, and task efficiency, aligning more closely with a straightforward, “tool”-oriented design. Importantly, intermediate profiles (\textit{Analytical Thinkers} \& \textit{Empathetic Sensors}) exhibited hybrid preferences, selectively drawing from both orientations. This finding challenges binary design assumptions and suggests that personality operates along continua rather than discrete categories. Their context-dependent proposal also indicates that even individuals with stable personality orientations may seek different support based on task demands, effort tolerance, or writing context.

These findings suggest that personalization in AI writing companions is not a matter of replacing shared functionality with segmented designs, but of layering variation on top of common foundations. Generic solutions risk overlooking meaningful differences in interaction expectations, while overly rigid personalization risks constraining users whose needs shift across contexts \cite{batool2025evaluating, alslaity2023panoramic}. Participatory design processes are therefore particularly valuable for surfacing these nuanced dimensions of fit, which may remain invisible in backend-only personalization approaches \cite{sanders2002user, zytko2022participatory, bach2024systematic}. Grounded in personality-informed co-design, our work offers both conceptual framing and practical guidance for designing more inclusive, adaptive, and commercially viable AI writing companions.

\subsubsection{Design Implications: Envisioning Adaptive Personalization}
The coexistence of convergence and divergence suggests that effective personalization in AI writing companions is less about fixed categorization and more about enabling users to arrive at configurations that feel appropriate to their current goals and contexts. These findings point toward a \textbf{layered personalization approach}:
\begin{itemize}
    \item Layer 1 - Establish fundamental capabilities as universal core features for all users.
    \item Layer 2 - Provide personality-aligned default configurations that reflect different writing orientations.
    \item Layer 3 - Enable contextual shifts in interaction style to accommodate changing tasks and writing situations.
\end{itemize}

This framing opens space for speculative design directions toward more adaptive AI writing companions:
\begin{itemize}
    \item \textbf{Pre-defined Modes:} Similar to light/dark modes, systems could offer distinct, selectable modes that reflect different cognitive and interpersonal orientations (e.g., structured, exploratory, supportive). Such modes offer simple, intuitive, and accessible entry points but may limit granularity in addressing needs.
    \item \textbf{Widget-Based Customization:} Instead of locking users into modes, systems can surface all features for diverse needs, as seen in many current applications. However, this approach may risk a steep learning curve and information overload. Alternatively, a modular, widget-based architecture could allow users to selectively assemble features based on task, mood, or workflow. Progressive disclosure and the ability to hide unused features may help balance flexibility with usability and allow personalization to evolve over time, fostering a stronger sense of ownership and agency \cite{wongso2024user, gero2019metaphoria, biermann2022tool}.
\end{itemize}

Future work could examine long-term adaptation trajectories, modular interface strategies, and AI writing companions that dynamically respond to users’ shifting preferences and contexts.

\subsection{The Role of Personality Match in Human-AI Teaming}
By observing participants' divergent responses to contrasting AI writing companions, our findings add to growing evidence that personality alignment plays an important role in human-AI teaming (e.g., \cite{wang2024ai, wang2020human, reza2025co}. Prior research has shown that alignment between team members (whether human or AI) influences collaboration performance and satisfaction \cite{salas2008teams, andrews2023role}. In writing, mismatches between users and system designs can create friction. This was evident in the mutual discomfort: \textit{Creative Feelers} expressed aversion to the rigid, structured, and logic-oriented “Solution Master,” while \textit{Practical Logicians} found the emotionally expressive “Empowering Pal” distracting and superficial. These mismatches shaped not only preference, but also potentially how participants engage with writing tasks (e.g., encourage idea exploration or cause unnecessary interruptions), undermining comfort, trust, and sustained engagement \cite{zhang2021ideal, huang2019human}. 
One might question whether reported fit reflected personality alignment or simply preference for a less cognitively demanding interface. However, the divergent preference patterns argue against this interpretation: Practical Logicians appreciated TSM's functional complexity as empowering, while Creative Feelers found the same complexity misaligned with their needs. This suggests that personality orientation meaningfully shaped how interface complexity was perceived and valued, rather than cognitive demand alone driving fit judgments.

Yet our results caution against interpreting personality matching as deterministic. The hybrid preferences exhibited by intermediate profiles, along with consistent endorsement for universal features, indicate that effective matching involves both baseline compatibility and flexible adaptation. Rather than assigning users to fixed AI personas based on personality assessment, personality may be better understood as an initial orientation that informs defaults (e.g., communication styles, feature prominence) while leaving room for users to customize and shift modes as needed. This approach respects personality as influential without treating it as prescriptive.

This dynamic view of interpersonal fit is particularly relevant in domains like writing, where motivation, emotional flow, and psychological safety are central \cite{kark2009alive, bruning2000developing}. Although we did not directly measure writing outcomes, participants frequently described changes in how they approached writing if interacting with a companion that “resonated” with them, such as greater willingness to iterate, smoother engagement with the writing process, and reduced procrastination. Future work could build on these observations by examining how design alignment affects measurable outcomes such as idea generation, writing quality, or goal completion.

Beyond individual writing, these insights extend to broader team-based settings. In multi-party interactions, AI agents capable of adapting to group dynamics can facilitate discussions, encourage participation, help moderate the group climate, and balance the personalities of team members \cite{zheng2022ux}. Research on team composition shows that personality makeup affects communication effectiveness and interpersonal harmony \cite{barrick1998relating, peeters2006personality}. Accordingly, attuning AI systems to both individual and collective preferences may help them evolve from tools to trusted collaborators and foster more respectful, cohesive, and productive collaboration {\cite{rezwana2022understanding, mcgrath2025collaborative, hemmer2025complementarity}}. While algorithmic performance remains essential, external interpersonal "fit" is equally vital in fostering harmony and team performance. We therefore advocate for adaptive AI systems that move beyond static, generalized solutions toward user-driven design considerations that complement algorithmic optimization, thus enabling more holistic forms of collaboration \cite{duin2023co, wu2025negotiating}.

\section{LIMITATIONS AND FUTURE WORK}
Our study has some limitations that open avenues for future research. First, to maintain conceptual clarity and avoid complex groupings, we focused only on Openness and Agreeableness when constructing writer profiles. Although theoretically motivated, this choice necessarily abstracts away other personality dimensions that may shape writing and collaboration. Future work could explore how additional Big Five traits interact with writing support. Pre-grouping participants may also introduce biases, as talkative individuals could steer discussions and obscure subtle personality differences. The sample size in the user review may further limit feedback diversity.

Second, the participant pool consisted solely of university students, making the findings most applicable to higher education and academic writing contexts (e.g., coursework, research papers) and limiting the generalizability. Future work could include more contexts, such as professional reports or business communications, to strengthen real-world applicability and address broader needs.

Moreover, our user review prioritized prototype functions, leaving room for experimentation with avatars, colors, or communication styles. Although these were discovered and incorporated into designs, we did not evaluate their impact in depth. Future studies could assess those factors to optimize design choices for diverse groups.
Additionally, while consolidating four profiles into a bipolar model helped foreground contrasting design orientations, this decision necessarily simplified intermediate profiles' nuanced preferences. The two prototypes also reflected divergent design philosophies rather than strict feature symmetry, and equal feature counts do not guarantee comparable experiential impact. Future work could develop more granular prototypes with feature sets calibrated for both structural parallelism and experiential equivalence to enable fairer and more systematic comparisons. Future work could also more rigorously disentangle personality alignment from cognitive demand through counterbalanced designs or explicit cognitive load measures.

Finally, this study served as a proof-of-concept using design provocations rather than a fully deployed system. We did not directly measure writing outcomes or long-term behavioral change. A next step would be to implement and deploy such systems in real writing workflows to evaluate how our personalized writing supports influence user experience and writing performance.




\bibliographystyle{ACM-Reference-Format}
\bibliography{sample-base}


\appendix

\section{LLM Prompts for Executable Demo}
\label{appendix:prompts}
Below, we list the specific prompt templates activated in each demo.

\begin{figure}[h]
    \centering
    \includegraphics[width=1\linewidth]{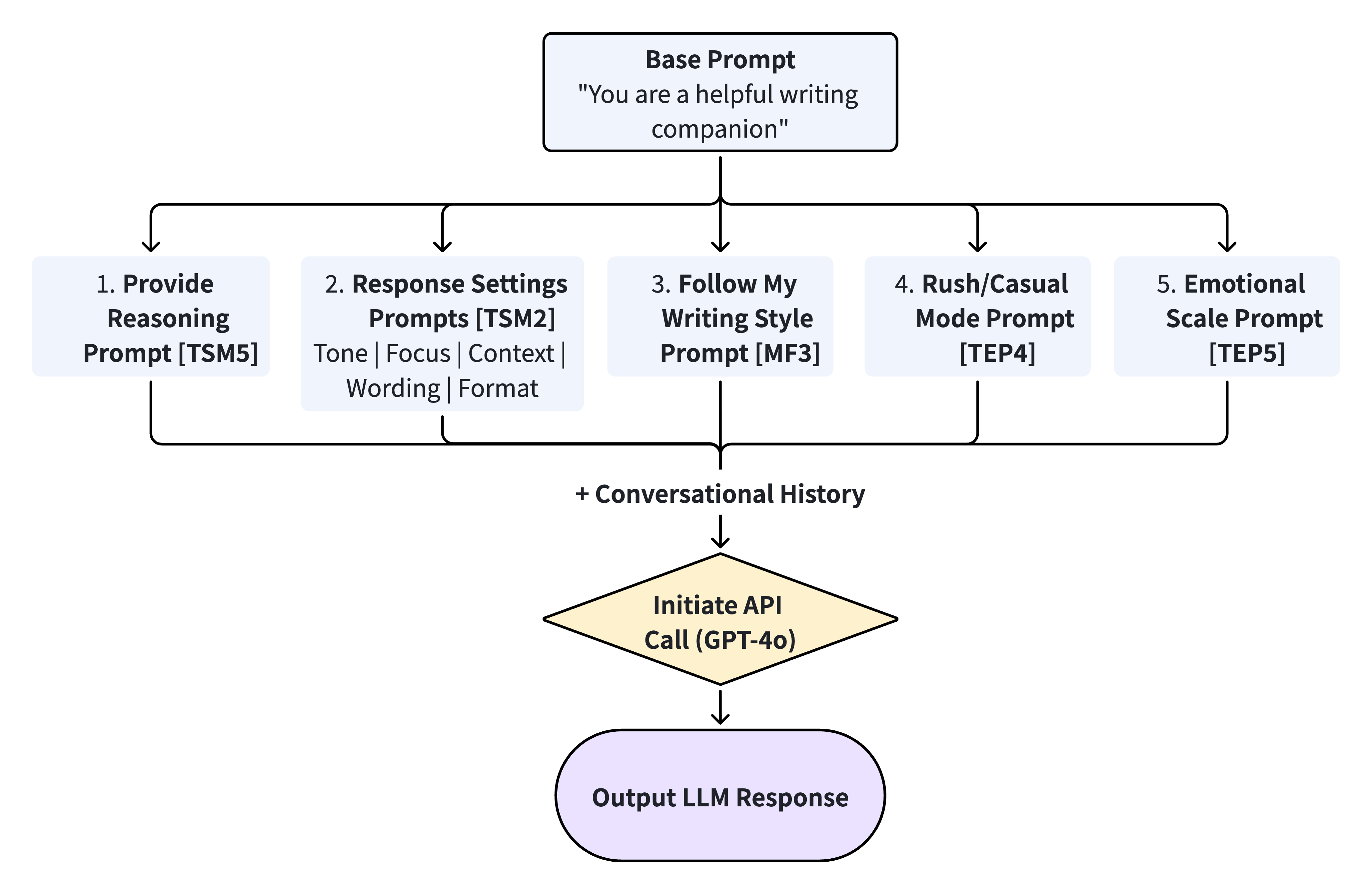}
    \caption{Prompt composition workflow. Base prompt branches into 5 parallel prompt components, then combined with conversation history and passed to the LLM API (GPT-4o) to generate the final response.}
    \Description{A flowchart showing the executable demo's prompt composition process. A base prompt branches into five parallel components — Provide Reasoning (TSM5), Response Settings (TSM2), Follow My Writing Style (MF3), Rush/Casual Mode (TEP4), and Emotional Scale (TEP5) — which combine with conversation history and are passed to GPT-4o to generate the final response.}
    \label{fig:prompts_workflow}
\end{figure}

\definecolor{opt}{HTML}{2563A0}
\definecolor{sechead}{HTML}{374151}

\newcommand{\SharedGrayBox}[2]{%
  \par\vspace{4pt}
  \noindent
  \begingroup
  \setlength{\fboxsep}{4pt}
  \setlength{\fboxrule}{0.5pt}
  \fcolorbox{gray!60}{white}{%
    \begin{minipage}{\linewidth}
      {\normalsize\bfseries\color{sechead} #1}\par
      \vspace{3pt}
      {\small #2\par}
    \end{minipage}
  }%
  \endgroup
  \par\vspace{3pt}
}

\SharedGrayBox{Shared Function - MF3: Follow My Writing Style}{%
  Analyze the provided writing sample and adopt the author's writing style, including their vocabulary, sentence structure, tone, and level of formality. Apply this style when generating or revising text.
}

\SharedGrayBox{The Empowering Pal - TEP4: Rush / Casual Mode}{%
  \textbf{\color{opt}Rush}: Keep responses short and direct. Use bullet points when possible. Skip pleasantries, filler words, and unnecessary elaboration. Lead with the answer or action, then add only essential context. Prioritize speed and clarity over completeness.\\[8pt]
  \textbf{\color{opt}Casual}: Respond in a relaxed, conversational tone as if chatting with a friend. Use everyday language, contractions, and a warm but brief style. Avoid overly formal structure or academic phrasing. Keep things light and approachable.%
}

\SharedGrayBox{The Empowering Pal - TEP5: Emotional Scale}{%
  \textbf{\color{opt}Harsh}: Be direct and critical. Point out flaws clearly without softening the language. Prioritize honest, unfiltered feedback over politeness.\\[6pt]
  \textbf{\color{opt}Aloof}: Stay detached and objective. Provide feedback matter-of-factly without emotional warmth or excessive encouragement.\\[6pt]
  \textbf{\color{opt}Neutral}: Balance honesty with politeness. Give constructive feedback in a straightforward but respectful manner.\\[6pt]
  \textbf{\color{opt}Friendly}: Be warm, supportive, and encouraging. Frame suggestions positively and acknowledge effort before offering improvements.\\[6pt]
  \textbf{\color{opt}Vibrant}: Be enthusiastic and highly encouraging. Celebrate strengths, use energetic language, frame all feedback as exciting opportunities for growth.%
}

\SharedGrayBox{The Solution Master - TSM2: Response Settings}{%
  \textbf{\color{sechead}Tone}\\[3pt]
  \hspace*{1em}\textbf{\color{opt}Neutral}: Maintain a neutral and professional tone.\\
  \hspace*{1em}\textbf{\color{opt}Playful}: Be a fun and creative assistant.\\
  \hspace*{1em}\textbf{\color{opt}Casual}: Keep the tone friendly and informal.\\
  \hspace*{1em}\textbf{\color{opt}Professional}: Use a formal and professional style.\\
  \hspace*{1em}\textbf{\color{opt}Formal}: Be very structured and formal.\\[3pt]
  \textbf{\color{sechead}Focus}\\[3pt]
  \hspace*{1em}\textbf{\color{opt}General}: Stick to general and balanced responses.\\
  \hspace*{1em}\textbf{\color{opt}Clarity}: Focus on providing clear and concise responses.\\
  \hspace*{1em}\textbf{\color{opt}Flow}: Prioritize a natural and smooth conversational flow.\\
  \hspace*{1em}\textbf{\color{opt}Grammar}: Emphasize grammatical accuracy and correctness.\\
  \hspace*{1em}\textbf{\color{opt}Empathy}: Respond with empathy and understanding.\\[3pt]
  \textbf{\color{sechead}Context}\\[3pt]
  \hspace*{1em}\textbf{\color{opt}General}: Stick to general and balanced responses.\\
  \hspace*{1em}\textbf{\color{opt}Academic}: Provide responses with an academic focus.\\
  \hspace*{1em}\textbf{\color{opt}Business}: Offer responses tailored for business needs.\\
  \hspace*{1em}\textbf{\color{opt}Casual}: Maintain a relaxed and informal tone.\\
  \hspace*{1em}\textbf{\color{opt}Creative}: Provide imaginative and creative responses.\\[3pt]
  \textbf{\color{sechead}Wording}\\[3pt]
  \hspace*{1em}\textbf{\color{opt}General}: Use general-purpose, understandable language.\\
  \hspace*{1em}\textbf{\color{opt}Kindergarten}: Use simple and child-friendly language.\\
  \hspace*{1em}\textbf{\color{opt}Secondary}: Use age-appropriate language for teenagers.\\
  \hspace*{1em}\textbf{\color{opt}University}: Adopt an academic style suitable for university students.\\
  \hspace*{1em}\textbf{\color{opt}Expert}: Use highly technical and expert-level language.\\[3pt]
  \textbf{\color{sechead}Format}\\[3pt]
  \hspace*{1em}\textbf{\color{opt}Current}: Offer responses of moderate length.\\
  \hspace*{1em}\textbf{\color{opt}Shortest}: Keep responses very brief.\\
  \hspace*{1em}\textbf{\color{opt}Shorter}: Provide concise answers.\\
  \hspace*{1em}\textbf{\color{opt}Longer}: Expand responses with more detail.\\
  \hspace*{1em}\textbf{\color{opt}Longest}: Give detailed and comprehensive responses.%
}

\SharedGrayBox{The Solution Master - TSM5: Provide Reasoning}{%
  For each major revision you make, include a brief rationale explaining why the change improves the text. Format each rationale in brackets immediately after the revised portion, e.g.\ \textbf{\color{opt}[Reason: ...]}. Focus on substance---clarity, accuracy, tone, or flow---not stylistic preference.
}

\section{Big Five Trait Distributions for Recruitment}
\label{appendix:stats}
\definecolor{LightGray}{gray}{0.9}
\newcolumntype{g}{>{\columncolor{LightGray}}c}

\begin{table*}
\centering
\Description{A table reporting Big Five personality trait scores across two groups. The upper section shows descriptive statistics (mean, Q1, median, Q3) for the screening pool (N=168) across five traits: Extraversion, Agreeableness, Conscientiousness, Neuroticism, and Openness. The lower section shows mean scores for the final sample (N=24) broken down by four writer profiles: Analytical Thinkers, Creative Feelers, Practical Logicians, and Empathetic Sensors. Gray-shaded columns indicate traits monitored but not used to define writer profiles.}
\begin{tabular}{
    >{\raggedright\arraybackslash}p{4cm}gcggc}
\toprule
 & \textbf{Extraversion} & \textbf{Agreeableness} & \textbf{Conscientiousness} & \textbf{Neuroticism} & \textbf{Openness} \\
\midrule

\multicolumn{6}{l}{\textbf{Screening Pool} ($N = 168$)} \\[2pt]
Mean        & 5.77 & 7.11 & 7.23 & 6.16 & 7.07 \\
Q1 (25\%)   & 4    & 6    & 6    & 5    & 6    \\
Median      & 6    & 7    & 7    & 6    & 7    \\
Q3 (75\%)   & 7    & 8    & 8    & 8    & 8    \\

\midrule
\multicolumn{6}{l}{\textbf{Final Sample by Writer Profile} ($N = 24$)} \\[2pt]
\textit{Analytical Thinkers}   & 6.83 & 6.00 & 8.00 & 4.00 & 7.83 \\
\textit{Creative Feelers}      & 5.67 & 8.50 & 5.50 & 7.50 & 7.83 \\
\textit{Practical Logicians}   & 6.50 & 6.67 & 7.00 & 5.17 & 4.50 \\
\textit{Empathetic Sensors}    & 6.50 & 7.83 & 7.17 & 5.83 & 6.17 \\

\bottomrule
\end{tabular}

\begin{flushleft}
\footnotesize
\textit{Note}: Means are reported for the final sample. Quartiles are computed from the screening pool, with writer profiles defined based on relative positioning along Openness and Agreeableness rather than strict threshold membership. Gray-shaded columns indicate Big Five dimensions monitored during recruitment but not used to define writer profiles.
\end{flushleft}
\end{table*}

\newpage
\section{Participant Profiles}
\label{appendix:profile}
\definecolor{LightGray}{gray}{0.9}
\newcolumntype{g}{>{\columncolor{LightGray}}c}

\begin{table*}
\centering
\Description{A table listing demographic and background information for all 24 participants organized by writer profile: Creative Feelers (P1–P6), Analytical Thinkers (P7–P12), Empathetic Sensors (P13–P18), and Practical Logicians (P19–P24). Columns include participant ID, Agreeableness score, Openness score, age, gender, education level, design experience, writing self-efficacy score, and whether they attended the user review session. A gray-shaded column indicates user review attendance.}
\begin{tabular}{cccclcccg}
\toprule
\textbf{P-ID} &
\textbf{Agreeableness} &
\textbf{Openness} &
\textbf{Age} &
\textbf{Gender} &
\textbf{Education} &
\textbf{\makecell{Design \\ Experience}} &
\textbf{\makecell{Writing \\ Self-Efficacy}} &
\textbf{\makecell{User \\ Review}} \\
\midrule

\multicolumn{9}{l}{\textbf{\textit{Creative Feelers}}} \\
\midrule
P1  & 9 & 9 & 24 & genderqueer & Senior & no   & 86.25  & x \\
P2  & 8 & 8 & 28 & male        & PhD    & some & 63.50  & x \\
P3  & 9 & 7 & 20 & female      & Junior & no   & 36.875 & x \\
P4  & 9 & 7 & 25 & male        & PhD    & some & 78.75  & \checkmark \\
P5  & 8 & 8 & 28 & male        & PhD    & no   & 49.375 & \checkmark \\
P6  & 8 & 8 & 26 & female      & Senior & no   & 85.00  & x \\

\midrule
\multicolumn{9}{l}{\textbf{\textit{Analytical Thinkers}}} \\
\midrule
P7  & 7 & 8 & 23 & female & Senior & yes  & 91.25  & x \\
P8  & 6 & 8 & 23 & female & Master & yes  & 91.25  & \checkmark \\
P9  & 7 & 8 & 28 & male   & PhD    & no   & 59.375 & \checkmark \\
P10 & 5 & 7 & 23 & male   & Master & yes  & 97.50  & x \\
P11 & 6 & 8 & 24 & female & Master & yes  & 50.00  & x \\
P12 & 5 & 8 & 21 & female & Senior & some & 81.25  & x \\

\midrule
\multicolumn{9}{l}{\textbf{\textit{Empathetic Sensors}}} \\
\midrule
P13 & 8 & 6 & 25 & female & Master    & yes  & 59.375 & \checkmark \\
P14 & 7 & 6 & 19 & female & Sophomore & no   & 70.625 & x \\
P15 & 8 & 6 & 20 & female & Senior    & some & 52.50  & \checkmark \\
P16 & 8 & 7 & 20 & male   & Sophomore & no   & 80.00  & x \\
P17 & 8 & 6 & 19 & female & Sophomore & no   & 90.00  & x \\
P18 & 8 & 6 & 25 & male   & Sophomore & yes  & 99.375 & x \\

\midrule
\multicolumn{9}{l}{\textbf{\textit{Practical Logicians}}} \\
\midrule
P19 & 7 & 4 & 21 & male   & Junior & yes & 88.13  & \checkmark \\
P20 & 7 & 4 & 31 & male   & Senior & no  & 70.63  & \checkmark \\
P21 & 6 & 3 & 20 & male   & Senior & yes & 67.50  & x \\
P22 & 7 & 6 & 20 & female & Junior & yes & 45.625 & x \\
P23 & 7 & 4 & 30 & female & Master & no  & 70.00  & x \\
P24 & 6 & 6 & 31 & female & PhD    & yes & 85.625 & x \\

\bottomrule
\end{tabular}

\begin{flushleft}
\footnotesize
\textit{Note.} Gray-shaded column indicates attendance for the user review session.
\end{flushleft}
\end{table*}


\end{document}